\documentclass{aa}
\usepackage[varg]{txfonts}   
\usepackage[english]{babel}
\usepackage{graphicx}
\usepackage{fancyvrb}
\usepackage[T1]{fontenc}
\usepackage[utf8]{inputenc}
\usepackage{amsmath}
\usepackage{amssymb}
\usepackage{amsfonts}
\usepackage{setspace}
\usepackage[nice]{units}
\usepackage{rotating}
\usepackage{natbib}
\usepackage[switch]{lineno}

\newcommand{\xmm}{\textit{XMM-Newton}}
\newcommand{\fermi}{\textit{Fermi}/LAT}
\newcommand{\swift}{\textit{Swift}}
\newcommand{\pks}{\mbox{PKS\,2004$-$447}}
\newcommand{\pmn}{\mbox{PMN\,J0948$+$0022}}
\newcommand{\pkss}{\mbox{PKS\,1502$+$036}}
\newcommand{\h}{\mbox{1H\,0323$+$342}}
\newcommand{\sbs}{\mbox{SBS\,0846$+$513}}

\newcommand{\kev}{\mathrm{keV}}

\title{The gamma-ray emitting radio-loud narrow-line Seyfert 1 galaxy \pks{}}
\titlerunning{The gamma-ray emitting radio-loud narrow-line Seyfert 1 galaxy \pks{}}
\subtitle{I. The X-ray View}
\author{A. Kreikenbohm\inst{\ref{inst1}, \ref{inst2}}
        \and R. Schulz\inst{\ref{inst1}, \ref{inst2}}
        \and M. Kadler \inst{\ref{inst1}}
        \and J. Wilms \inst{\ref{inst2}}
        \and A. Markowitz\inst{\ref{inst2}, \ref{inst3}, \ref{inst4}}
        \and C.S. Chang\inst{\ref{inst5}}
        \and B. Carpenter\inst{\ref{inst6}, \ref{inst7}}
        \and D. Els\"{a}sser\inst{\ref{inst1}}
        \and N. Gehrels\inst{\ref{inst6}}
        \and K. Mannheim\inst{\ref{inst1}}
        \and C. M\"{u}ller\inst{\ref{inst7}, \ref{inst1}}
        \and R. Ojha\inst{\ref{inst6}, \ref{inst8}, \ref{inst9}}
        \and E. Ros \inst{\ref{inst10}, \ref{inst11}, \ref{inst12}}
        \and J. Trüstedt\inst{\ref{inst1}}
}
\institute{Lehrstuhl f\"{u}r Astronomie, Universit\"{a}t Würzburg, Campus Hubland Nord, Emil-Fischer-Straße 31, 97074, W\"{u}rzburg, Germany \label{inst1}
\and Dr Karl Remeis-Observatory and Erlangen Centre for Astroparticle Physics, Sternwartstr. 7, 96049 Bamberg, Germany \label{inst2}
\and University of California, San Diego, CASS, 9500 Gilman Dr., MC~0424, La Jolla, CA 92093-0424, USA \label{inst3}
\and Alexander von Humboldt Fellow \label{inst4}
\and Joint ALMA Observatory, ESO, Santiago, Chile \label{inst5}
\and Astrophysics Science Division, NASA Goddard Space Flight Center, Greenbelt, MD 20771 USA  \label{inst6}
\and Department of Astrophysics/IMAPP, Radboud University Nijmegen, PO Box 9010, 6500 GL, Nijmengen, The Netherlands \label{inst7}
\and Catholic University of America, Washington, DC 20064, USA \label{inst8}
\and University of Maryland Baltimore County/CRESST, Baltimore, MD 21250, USA \label{inst9}
\and Max-Planck-Institut für Radioastronomie, Auf dem H\"{u}gel 69, 53121, Bonn, Germany \label{inst10}
\and Observatori Astron\`{o}mic, Univ. Val\`{e}ncia, 46980 Paterna Val\`{e}ncia Spain \label{inst11}
\and Dept. d'Astronomia i Astrof\'{i}sica, Univ. Val\`{e}ncia, 46100 Burjassot, Val\`{e}ncia, Spain \label{inst12}
}
\date{Received 15 August 2014 / Accepted 02 October 2015} 

\abstract{As part of the TANAMI multiwavelength progam, we discuss new X-ray observations of the $\gamma$-ray and radio-loud narrow line \mbox{Seyfert 1} galaxy ($\gamma$-NLS1) \pks{}. The active galaxy is a member of a small sample of radio-loud NLS1s detected in $\gamma$-rays by the \textit{Fermi} Large Area Telescope. It stands out for being the radio-loudest and the only southern-hemisphere source in this sample. We present results from our X-ray monitoring program comprised of \swift{} snapshot observations from 2012 through 2014 and two new X-ray observations with \xmm{} in 2012. Supplemented by archival data from 2004 and 2011, our data set allows for a careful analysis of the X-ray spectrum and variability of this peculiar source. The (0.5--10)\,keV spectrum is described well by a power law ($\Gamma\sim1.6$), which can be interpreted as non-thermal emission from a relativistic jet. The source exhibits moderate flux variability on timescales of both months and years. Correlated brightness variations in the (0.5--2)\,keV and (2--10)\,keV bands are explained by a single variable spectral component, such as the one from the jet. A possible soft excess seen in the data from 2004 cannot be confirmed by the new \xmm{} observations taken during low-flux states. 
Any contribution to the total flux in 2004 is less than 20\% of the power-law component. The (0.5--10)\,keV luminosities of \pks{} are in the range of (0.5--2.7)$\times10^{44}\,\mathrm{erg\,s}^{-1}$. A comparison of the X-ray properties among the known $\gamma$-NLS1 galaxies shows that in four out of five cases the X-ray spectrum is dominated by a flat power law without intrinsic absorption. These objects are moderately variable in their brightness, while spectral variability is observed in at least two sources. The major difference across the X-ray spectra of $\gamma$-NLS1s is the luminosity, which spans a range of almost two orders of magnitude from $10^{44}\,\mathrm{erg\,s}^{-1}$ to $10^{46}\,\mathrm{erg\,s}^{-1}$ in the (0.5--10)\,keV band.
}


\usepackage{color}

\begin{document}

\maketitle
\section{Introduction}
 \label{sec-Intro}
The recent detection of variable $\gamma$-ray emission from radio-loud narrow line Seyfert 1 galaxies ($\gamma$-NLS1 hereafter) suggests that there are relativistic jets in these sources, which are similar to other types of active galactic nuclei (AGN), i.e. blazars and radio galaxies
\citep{Abdo2009a,Foschini2013,Dammando2013b,Foschini2015}. Typical properties of radio-quiet NLS1s are black hole masses that range from $10^6\,M_\sun$ to $10^8\,M_\sun$ \citep[see, e.g.,][]{Hayashida2000,Wandel2000,Boroson2002}, an accretion rate close to the Eddington limit, and spiral host galaxies \citep[see, e.g.,][]{Foschini2012}. Their detection in $\gamma$-rays poses serious questions as to the well-known paradigm that such powerful jets are hosted by AGN in elliptical galaxies where a supermassive black hole (SMBH) of at least $10^{8}M_\sun$ accretes at a few percent of the Eddington rate \citep[see, e.g.,][for a review]{Marscher2009}. At the intersection of two rare types of AGN (NLS1 and radio-loud AGN), $\gamma$-NLS1s are unique laboratories for studying the jet formation in AGN. 

Formally, NLS1 galaxies are classified by their strong and narrow optical Balmer emission lines from the broad line region. The intensities of the narrow lines are strong with respect to the forbidden [$\ion{O}{iii}$] $\lambda$5007 line, i.e., FWHM($\ion{H}{}\beta)\le2000\,\mathrm{km\,s}^{-1}$ and [$\ion{O}{iii}$]/$\ion{H}{}\beta\le3$ \citep{Osterbrock1985}. These properties are typically accompanied by a strong $\ion{Fe}{ii}$ emission line complex \citep{Grupe2004} and strong characteristic high-ionization line emission from the broad line region. The origins of the strong $\ion{Fe}{ii}$ emission and of the narrow Balmer lines are still a topic of debate.
In X-rays, compared to other Seyfert 1 galaxies, NLS1 exhibit typically stronger X-ray flux variability on short timescales \citep[see, e.g.,][]{VaughanFabian2003,Grupe2004,Kara2013} and a strong soft excess below $2\,\kev$ \citep{Vaughan1999, Grupe2004}. 

Among NLS1s, only 7\% are radio-loud \citep{Komossa2006,Yuan2008}, of which only a small number have been detected in $\gamma$-rays \citep{Abdo2009a,Foschini2013,Dammando2013b}. Multiwavelength campaigns have been performed on three of the known $\gamma$-NLS1s detected by the \textit{Fermi} Large Area Telescope (LAT; \citealt{Atwood2009}): \pmn{} \citep{Abdo2009a}, \h{} \citep{Paliya2014}, and \sbs{} \citep{Dammando2013b}. 
We present results from a multiwavelength monitoring of the radio-loudest source from this elusive $\gamma$-NLS1 sample.

The active galaxy \pks{} is a $\gamma$-NLS1 galaxy at a redshift of $z=0.24$ \citep{Drinkwater1997} with a radio loudness parameter $R=f_{4.9\,\mathrm{GHz}}$/$f_B\ge 1700$ (\citealt[][hereafter Osh01]{Kellermann1989,Oshlack2001}; \citealt[][hereafter G06]{Gallo2006}). Its optical properties are consistent with the NLS1 classification: \mbox{a flux ratio [$\ion{O}{iii}$]/$\ion{H}{}\beta=1.6$} and \mbox{FWHM$(\mathrm{H}\beta)=1447\,\mathrm{km\,s}^{-1}$}, even though the $\ion{Fe}{ii}$ emission (\mbox{$EW_{\ion{Fe}{ii}}\le10\AA$}) is rather weak (Osh01).
Based on the empirical relation between BLR size and optical luminosity \citep{Kaspi2000}, Osh01 estimated the black hole mass from the $\ion{H}{}\beta$ line to be $10^{\,6.7}\,M_\sun$, using high-resolution data from the Siding Spring 2.3 m Telescope. Studies in the X-rays \citep[see e.g., G06;][]{Paliya2013} suggest an unobscured spectrum in the (0.3--10)\,keV band, a possible soft excess below 1\,keV, and mild long-term flux variability. 

We observed this source as part of the TANAMI multiwavelength monitoring program \citep{Ohja2010,Kadler2015}.
Here we present results from the analysis of the X-ray spectra. In addition to archival data from 2004, we use two new \xmm{} pointing observations performed in 2012 to explore the source's long-term X-ray spectral variability. For the study of X-ray flux variability, \pks{} has been monitored with \swift{} since 2011.

In a second paper (Schulz et al. submitted, Paper II), we report on the radio properties of \pks{}, including the first TANAMI VLBI image at 8.4\,GHz. VLBI imaging shows an elongated, one-sided radio jet extending from a bright compact core. Simultaneous multifrequency observations with the Australia Telescope Compact Array (ATCA) over four years reveal a persistent steep spectrum. The estimated large-scale size of the radio emission and the spectrum are consistent with compact steep spectrum (CSS) sources, which are generally considered to be young radio galaxies \citep{Odea1998}. 

The paper is organized as follows. In Sect. \ref{sec-obs}, the monitoring program and data reduction methods are described. The analysis of X-ray data and its results are reported in Sect. \ref{sec-ana} and discussed in Sect. \ref{sec-ana-dis}, which also includes a comparison with X-ray properties of other $\gamma$-NLS1s, CSS galaxies, and blazars. The main results of the work are summarized in Sect. \ref{sec-sum}.

\section{Observations and data reduction}
\label{sec-obs}
\pks{} has been monitored in X-rays since May 2013 as part of the TANAMI multiwavelength program. We used the space observatories \xmm{} \citep{Jansen2001} and \swift{} \citep{Gehrels2004}, which  provide X-ray data  and optical/UV coverage. High-angular-resolution radio observations have been performed since  October 2010 by the TANAMI VLBI program. In addition, the source has been observed several times by ATCA. Simultaneous $\gamma$-ray data are provided by the \fermi{}. 
The X-ray monitoring consists of two \xmm{}{} pointings in 2012 separated by five months and connected by four \swift{} observations. We have continued observations with \swift{} in 2013 and 2014. To study the long-term behavior of \pks{}, archival data from \xmm{}{} and \swift{} prior to 2013 May have been included in the analysis. 
Detailed information on all observations is provided in Table \ref{tab-ObsInfo}. 

\begin{table*}
\caption{Details of all \xmm{}{} and \swift{} observations of \pks{}.}
\label{tab-ObsInfo}
\centering
\begin{tabular}{llrrr}
\hline \hline
Inst-ObsDate &ObsId &  Duration &  Net exposure & Spectrum counts  \\
 &  & [ks]  & [ks]&[counts]  \\
(1) & (2) & (3) & (4) & (5) \\
\hline
X-2012-10-18  &0694530201                       & 39.8  &34.5   &        10788\\
X-2012-05-01  &0694530101                       & 37.9  &36.5   &        6125\\
X-2004-04-11  &0200360201                       & 41.9  &38.4   &        15831\\
\hline
S-2014-03-14  &000324920[16,17]\tablefootmark{a}& 17.2 &17.1    &        248\\ 
S-2013-11-19  &000324920[14,15]\tablefootmark{b}& 17.7 &16.7    &        523\\ 
S-2013-10-13  &0003249200[09-13]\tablefootmark{c}& 16.5         &16.5   &        302\\
S-2013-09-27  &00032492007                      & 8.3   &8.3    &        234\\
S-2013-07-07  &0003249200[5,6]\tablefootmark{d} & 23.1  &23.0   &        444\\
S-2012-09-30  &00032492004                      & 6.0   &6.0    &        63\\
S-2012-09-12  &00032492003                      & 5.4   &5.4    &        63\\
S-2012-07-22  &00032492002\tablefootmark{e}     & 2.3   &-      &        -\\
S-2012-07-03  &00032492001                      & 4.9   &4.9    &        67\\
S-2012-03-14  &00091031007                      & 7.3   &7.3    &        60\\
S-2011-11-15  &00091031006                      & 7.1   &7.1    &        73\\
S-2011-09-17  &00091031005                      & 7.5   &7.34   &        167\\
S-2011-07-29  &00091031004\tablefootmark{e}     & 1.6   &-      &        -\\
S-2011-07-25  &00091031003\tablefootmark{e}     & 0.6   &-      &        -\\
S-2011-07-14  &00091031002                      & 5.0   &4.9    &        56\\ 
S-2011-05-15  &00091031001                      & 6.8   &6.8    &        61\\
\hline
\end{tabular}
\tablefoot{(1) Instrument and observation date. We use the format Inst-yyyy-mm-dd: X - \xmm{} EPIC, S - \swift{} XRT. (2) Observation identifier. (3) Observation duration. (4) Net exposure time after screening for flaring particle background. (5) Background-subtracted total number of counts in the (0.3--10)\,keV band. \textbf{Merged observations:} 
\tablefoottext{a} {00032492016 (2014-03-14, 7.6\,ks) and 00032492017 (2014-03-16, 9.7\,ks).}
\tablefoottext{b} {00032492014 (2013-11-19, 5.5\,ks) and 00032492015 (2013-11-20, 12.2\,ks).}
\tablefoottext{c} {00032492009 (2013-10-13, 4.7\,ks), 00032492010 (2013-10-20, 4.11\,ks), 00032492011 (2013-10-27, 2.35\,ks), 00032492012 (2013-10-29, 1.51\,ks), and 00032492013 (2013-11-03, 3.88\,ks).}
\tablefoottext{d} {00032492005 (2013-07-07, 11.4\,ks) and 00032492006 (2013-07-14, 11.6\,ks).}
\tablefoottext{e} {Because of the relatively short exposure time, this observation was not included in the analysis.}
}
\end{table*}

\subsection{\xmm{} observations} 
\label{sec-obs.1}
We observed \pks{} twice with \xmm{} in 2012: for 37.9\,ks on 2012 May 1 and for 39.9\,ks on 2012 October 18, hereafter referred to as X-2012-05-01 and X-2012-10-18. The archival \xmm{} observation from 2004 April 11 (for 41.9\,ks, X-2004-04-11 hereafter) has been discussed in detail by G06. All observations were performed with the European Photon Imaging Camera (EPIC) using PN \citep{Strueder2001} and MOS \citep{Turner2001esa} CCD arrays and the Reflecting Grating Spectrometer (RGS, \citealt{denherder2001}), as well as the Optical Monitor (OM, \citealt{Mason2001}). The EPIC operated in full-frame mode with the medium filter in 2004 and the thin filter in 2012.
Observation data files (ODFs) were processed to create calibrated event lists and full frame images using standard methods with the \xmm{}{} Science Analysis System (SAS Version 12). Lightcurves were extracted for the $(0.5-2)\,\kev$, $(2-10)\,\kev$, and $(0.5-10)\,\kev$ energy ranges, adopting a circular extraction region with 30 arcsec radius for the source. The background regions were chosen from a source-free circular region of the same radius on the same chip with similar distances to the readout nodes to ensure similar background noise levels. Screening for flaring particle background was done following the method to maximize the signal-to-noise ratio (S/N) by \citet{Piconcelli2004}. Pile-up was negligible in all observations. The net exposure times are listed in Table \ref{tab-ObsInfo}. Source and background spectra were extracted for single and double event patterns from the filtered event lists. Response files were generated using  \textsc{rmfgen} and \textsc{arfgen} of the SAS software package. The RGS data had insufficient 
counts for a dedicated spectral analysis.

\subsection{\swift{} observations} 
\label{sec-obs.2}
We monitored \pks{} with the \swift{} X-ray telescope (XRT, \citealt{Burrows2005}) from the beginning of 2012 through the beginning of 2014. In addition, we used archival data from 2011, which were discussed by \cite{Paliya2013}. Details of each observation are listed in \mbox{Table \ref{tab-ObsInfo}}. Three observations had relatively short exposures (2.3\,ks, 1.6\,ks, and 0.6\,ks). Owing to the low S/N in the respective X-ray spectra, they are not considered in our analysis. For all other observations, we used data from the XRT, which operated in photon-counting mode (PC). The data were cleaned, and calibrated event files were created using the standard filtering methods and the \textsc{xrtpipeline} task, distributed in the HEASARC (v6.12.0) within the HEASoft package. Source and background spectra and lightcurves in the $(0.5-10)\,\mathrm{keV}$ energy range were created, using a circular source region with a radius of 35 arcsec. The net exposure times are also listed in \mbox{Table \ref{tab-ObsInfo}}. To increase the 
S/N of the spectra, \swift\  observations that were observed within a few days of each other were merged after checking that the spectra did not show significant flux or spectral variability between the averaged epochs.

\section{X-ray analysis}
\label{sec-ana}
Spectral fitting was done using the \textit{\emph{Interactive Spectral Interpretation System}} 
(ISIS, Version 1.6.2-30, \citealt{Houck2000}). Unless stated otherwise, uncertainties correspond to 90\% confidence limits on one parameter of interest \mbox{($\Delta\chi^2=2.7$)}. In the following, the standard cosmological model (\mbox{$H_0=70\,\mathrm{km\,s}^{-1}\,\mathrm{Mpc}^{-1}$}, $\Omega_\mathrm{M}=0.3$, and $\Lambda=0.7$) is assumed. 

\subsection{\xmm{}{} PN/MOS data}
\label{sec-ana-xmm}

\begin{table*}
\caption{Best-fit results of the absorbed power-law model for \xmm{} and \swift{} spectra. }
\label{tab-bfpar-pow}
\centering
\begin{tabular}{llcccccr}
\hline \hline
Inst-ObsDate & ObsID&  $\Gamma$ & $F_{0.5-10\kev}$ & $F_{0.5-2\kev}$ &  $F_{2-10\kev}$  & $L_{0.5-10\kev}$ & stat/dof (statistic)\\    
(1) & (2) & (3) & (4) & (5) & (6) & (7) & (8) \\
\hline
X-2012-10-18 & 0694530201 & $1.61^{+0.03}_{-0.02}$ & $0.70^{+0.02}_{-0.02}$ & $0.23^{+0.01}_{-0.01}$ & $0.48^{+0.03}_{-0.02}$ & $1.11\pm 0.03$ & 296.4/258 ($\chi^2$)\\ 
X-2012-05-01 & 0694530101 & $1.56^{+0.05}_{-0.05}$ & $0.48^{+0.03}_{-0.02}$ & $0.15^{+0.01}_{-0.00}$ & $0.35^{+0.04}_{-0.02}$ & $0.75\pm 0.04$ & 135.3/131 ($\chi^2$)\\ 
X-2004-04-11 & 0200360201 & $1.52^{+0.02}_{-0.02}$ & $1.45^{+0.03}_{-0.03}$ & $0.43^{+0.01}_{-0.01}$ & $1.06^{+0.04}_{-0.04}$ & $2.26\pm 0.05$ & 388.3/398 ($\chi^2$)\\ 
\hline
S-2014-03-14 & 000324920[16,17]$^{l}$ & $1.43^{+0.15}_{-0.12}$ & $0.95^{+0.13}_{-0.15}$ & $0.26^{+0.03}_{-0.03}$ & $0.75^{+0.16}_{-0.20}$ & $1.45\pm 0.23$ & 622.1/948  ($C$) \\ 
S-2013-11-19 & 000324920[14,15]$^{h}$ & $1.64^{+0.12}_{-0.12}$ & $1.52^{+0.14}_{-0.16}$ & $0.51^{+0.05}_{-0.05}$ & $0.91^{+0.14}_{-0.17}$ & $2.44\pm 0.26$ & 594.7/948  ($C$)\\ 
S-2013-10-13 & 000324920[09-13]$^{m}$ & $1.52^{+0.15}_{-0.13}$ & $0.99^{+0.13}_{-0.14}$ & $0.30^{+0.04}_{-0.04}$ & $0.71^{+0.15}_{-0.19}$ & $1.55\pm 0.23$ & 517.2/948  ($C$)\\ 
S-2013-09-27 & 00032492007$^{h}$ & $1.47^{+0.17}_{-0.18}$ & $1.54^{+0.22}_{-0.26}$ & $0.43^{+0.06}_{-0.07}$ & $1.21^{+0.28}_{-0.36}$ & $2.38\pm 0.42$ & 481.8/948  ($C$)\\ 
S-2013-07-07 & 0003249200[5,6]$^{m}$ & $1.41^{+0.12}_{-0.09}$ & $1.16^{+0.12}_{-0.14}$ & $0.31^{+0.03}_{-0.03}$ & $0.92^{+0.16}_{-0.19}$ & $1.77\pm 0.22$ & 671.6/948  ($C$)\\ 
S-2012-09-30 & 00032492004$^{l}$ & $1.63^{+0.25}_{-0.28}$ & $0.52^{+0.13}_{-0.17}$ & $0.16^{+0.04}_{-0.05}$ & $0.26^{+0.10}_{-0.15}$ & $0.82\pm 0.25$ & 244.8/948  ($C$)\\ 
S-2012-09-12 & 00032492003$^{l}$ & $1.77^{+0.32}_{-0.32}$ & $0.56^{+0.13}_{-0.17}$ & $0.22^{+0.05}_{-0.06}$ & $0.41^{+0.18}_{-0.30}$ & $0.93\pm 0.26$ & 270.2/948  ($C$)\\ 
S-2012-07-03 & 00032492001$^{l}$ & $1.75^{+0.32}_{-0.33}$ & $0.63^{+0.15}_{-0.20}$ & $0.24^{+0.06}_{-0.07}$ & $0.50^{+0.21}_{-0.35}$ & $1.03\pm 0.29$ & 249.8/948  ($C$)\\ 
S-2012-03-14 & 00091031007$^{l}$ & $1.57^{+0.24}_{-0.27}$ & $0.42^{+0.11}_{-0.14}$ & $0.13^{+0.03}_{-0.04}$ & $0.31^{+0.12}_{-0.19}$ & $0.67\pm 0.21$ & 271.9/948  ($C$)\\ 
S-2011-11-15 & 00091031006$^{l}$ & $2.04^{+0.33}_{-0.31}$ & $0.39^{+0.09}_{-0.11}$ & $0.18^{+0.04}_{-0.05}$ & $0.19^{+0.08}_{-0.14}$ & $0.68\pm 0.17$ & 243.5/948  ($C$)\\ 
S-2011-09-17 & 00091031005$^{m}$ & $1.45^{+0.19}_{-0.20}$ & $1.18^{+0.20}_{-0.24}$ & $0.33^{+0.05}_{-0.06}$ & $0.79^{+0.21}_{-0.28}$ & $1.82\pm 0.37$ & 405.2/948  ($C$)\\ 
S-2011-07-14 & 00091031002$^{l}$ & $1.66^{+0.34}_{-0.34}$ & $0.63^{+0.16}_{-0.22}$ & $0.21^{+0.05}_{-0.07}$ & $0.49^{+0.18}_{-0.30}$ & $1.02\pm 0.32$ & 258.2/948  ($C$)\\ 
S-2011-05-15 & 00091031001$^{l}$ & $1.38^{+0.23}_{-0.26}$ & $0.54^{+0.14}_{-0.19}$ & $0.14^{+0.04}_{-0.05}$ & $0.56^{+0.21}_{-0.34}$ & $0.82\pm 0.27$ & 273.2/948  ($C$)\\ 
\hline 
S-low           & - & $1.53^{+0.12}_{-0.09}$ & $0.53^{+0.05}_{-0.08}$ & $0.16^{+0.01}_{-0.02}$ & $0.33^{+0.07}_{-0.12}$ &  $0.84\pm0.11$  &  15.7/12 ($\chi^2$)\\ 
S-medium        & - & $1.50^{+0.10}_{-0.10}$ & $1.02^{+0.07}_{-0.12}$ & $0.30^{+0.02}_{-0.03}$ & $0.70^{+0.09}_{-0.14}$ &  $1.59\pm0.16$  &  19.7/27 ($\chi^2$)\\ 
S-high          & - & $1.57^{+0.12}_{-0.12}$ & $1.49^{+0.10}_{-0.22}$ & $0.46^{+0.03}_{-0.06}$ & $0.94^{+0.15}_{-0.25}$ &  $2.36\pm0.28$  &  22.6/23 ($\chi^2$)\\ 
\hline
\end{tabular}
\tablefoot{Best-fit parameters for the absorbed power law. The absorption is fixed to its Galactic absorption, $3.17\times 10^{20}\,\mathrm{cm}^{-2}$, based on the survey by \cite{Kalberla2005}. (1) Instrument and observation dates as in Table \ref{tab-ObsInfo}. (2) Observation ID, [n,m] indicates merged spectra. (3) Power-law photon index. (4,5,6) Absorption-corrected fluxes in units of $10^{-12}\,\mathrm{erg\,cm}^{-2}\,\mathrm{s}^{-1}$ for the $(0.5-10)\,\kev$, $(0.5-2)\,\kev$, $(2-10)\,\kev$ energy ranges. (7) Absorption-corrected luminosity in units of $10^{44}\,\mathrm{erg\,s}^{-1}$. (8) Statistic value per degrees of freedom (dof) for the best fit to the full band. The \swift{} data are not rebinned. Since the degrees of freedom are then given by the energy grid of the XRT it is constant for all \swift{} observations. \tablefoottext{l,m,h} {Observations included in the merged \swift{} data set of low (l), medium (m), or high (h) fluxes (see \mbox{Sects. \ref{sec-ana-xmm} \& \ref{sec-ana-xrt}} for details).}}
\end{table*}

\begin{figure}
\centering
  \begin{minipage}{0.95\linewidth}
  \resizebox{\hsize}{!}{\includegraphics[width=\linewidth]{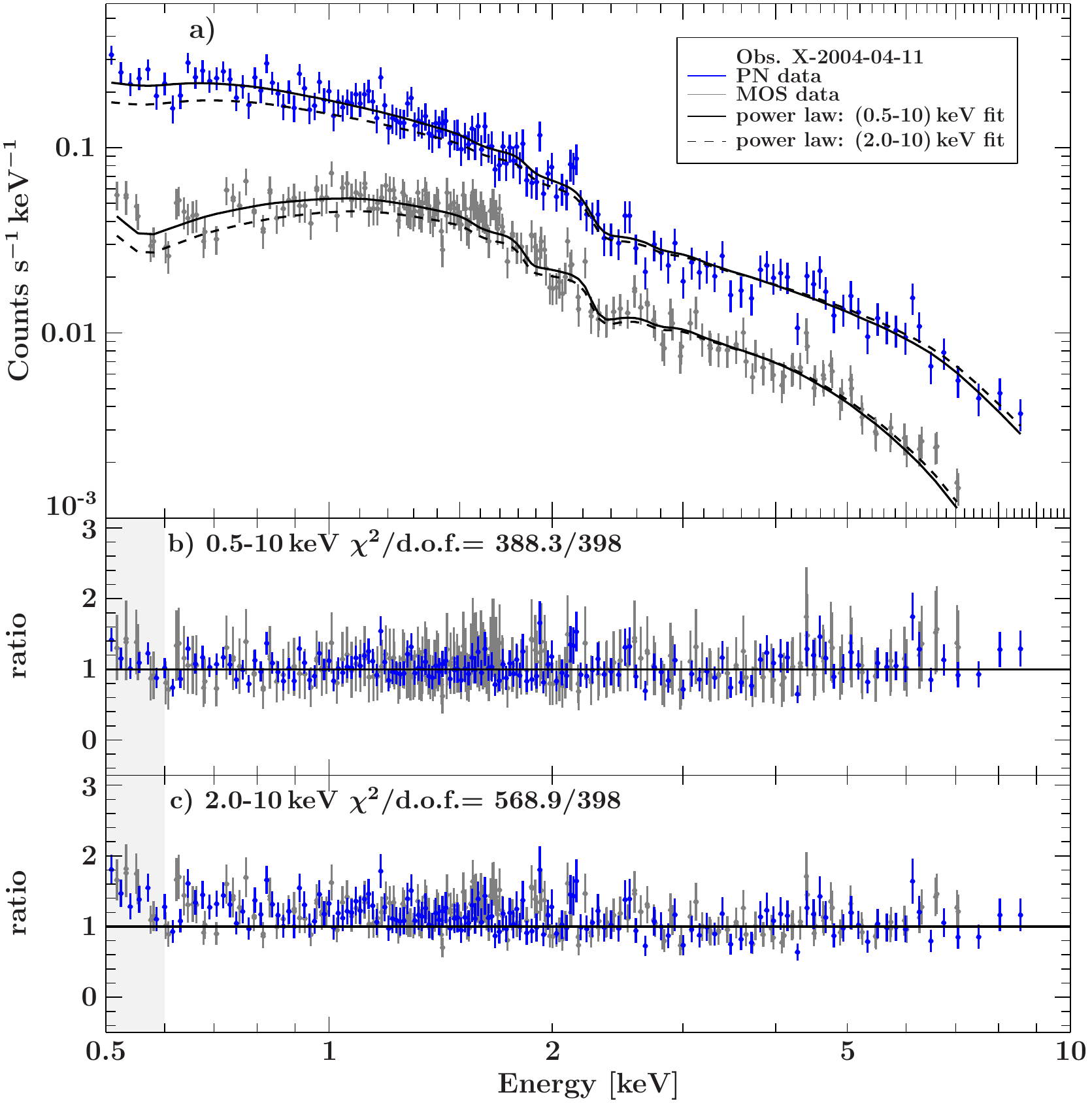}} 
  \end{minipage}
  \begin{minipage}{0.95\linewidth}
   \resizebox{\hsize}{!}{\includegraphics[width=\linewidth]{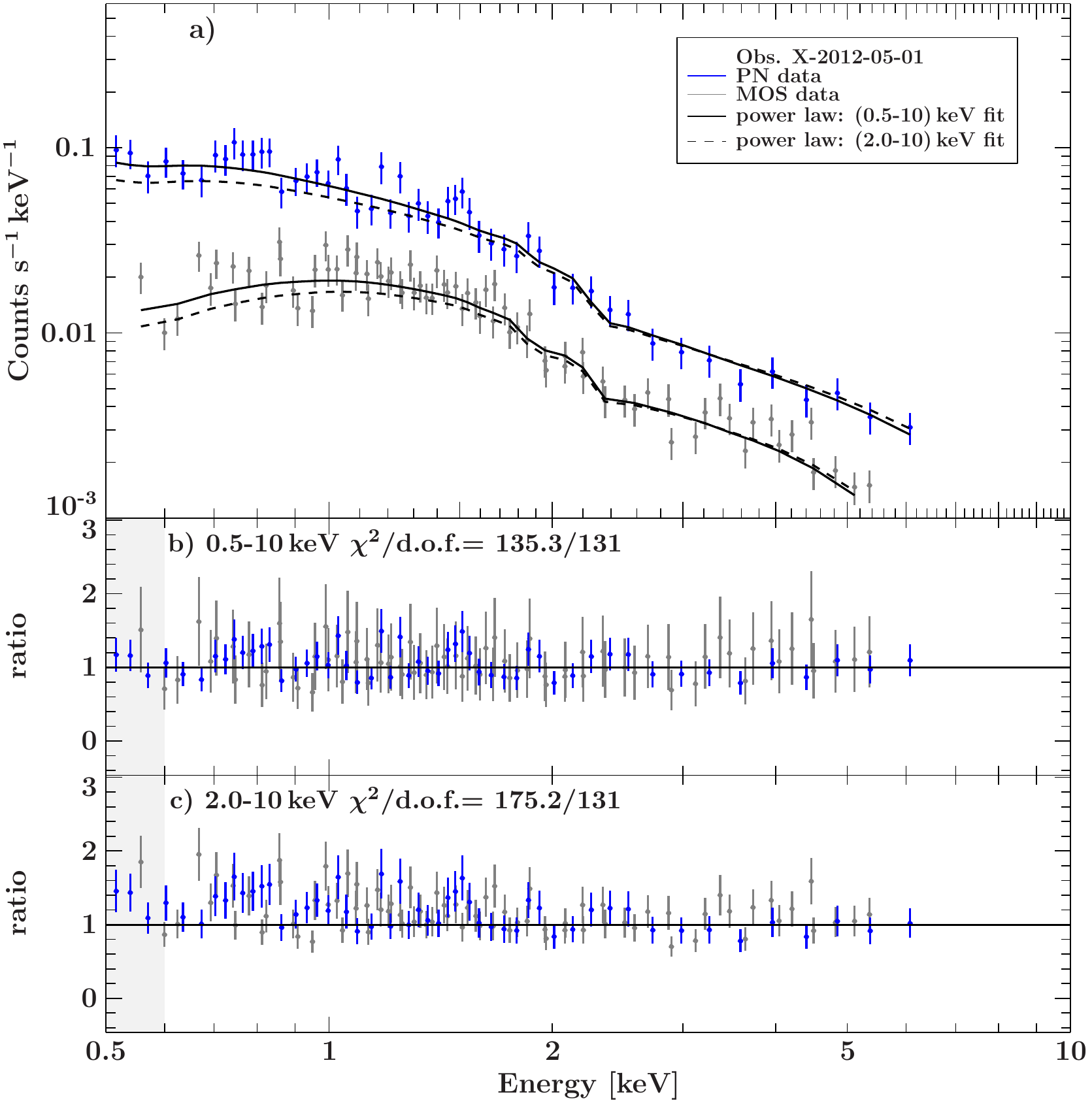}} 
  \end{minipage}
  \caption{ \xmm{} spectrum and best-fit results of X-2004-04-11 (top) and X-2012-05-01 (bottom). \textit{Upper panel a):} EPIC pn (blue) and MOS (gray) data, as well as corresponding best fits for an absorbed power law evaluated over (0.5--10)\,keV (black solid line) and (2--10)\,keV (black dashed line). \textit{Mid panel b):} Data-to-model ratio for the best fit over the whole (0.5--10)\,keV energy range. \textit{Lower panel c):} Respective residuals for fitting the (2--10)\,keV energy range alone, extrapolated to lower energies. A tentative soft excess below 0.6\,keV is only observed for X-2004-04-11 (compare shaded regions).}  
\label{fig-xmm_abspow02}
\end{figure}
\begin{figure}
\centering
\begin{minipage}{0.95\linewidth}
   \resizebox{\hsize}{!}{\includegraphics[width=\linewidth]{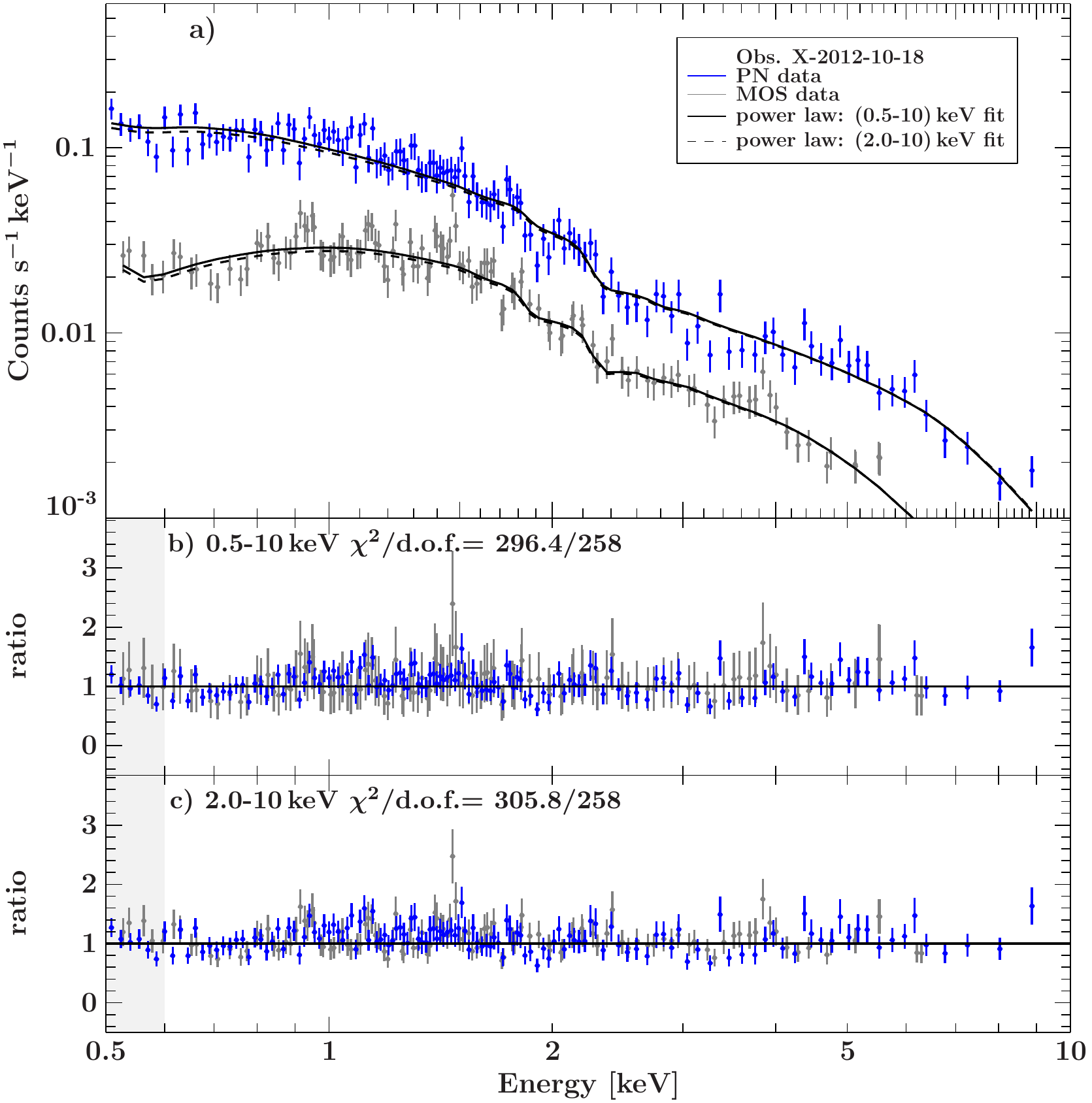}} 
     \end{minipage}
   \caption{ \xmm{} spectrum and best-fit results of X-2012-10-18. \textit{Upper panel a):} EPIC pn (blue) and MOS (gray) data, as well as corresponding best fits for an absorbed power law evaluated over (0.5--10)\,keV (black solid line) and (2--10)\,keV (black dashed line). \textit{Mid panel b):} Data-to-model ratio for the best fit over the whole (0.5--10)\,keV energy range. \textit{Lower panel c):} Respective residuals for fitting the (2--10)\,keV energy range alone, extrapolated to lower energies. A tentative soft excess below 0.6\,keV is only observed for X-2004-04-11 (compare shaded regions).}
\label{fig-xmm_abspow06}
\end{figure}

\begin{figure}
\centering
   \resizebox{\hsize}{!}{\includegraphics[width=\linewidth]{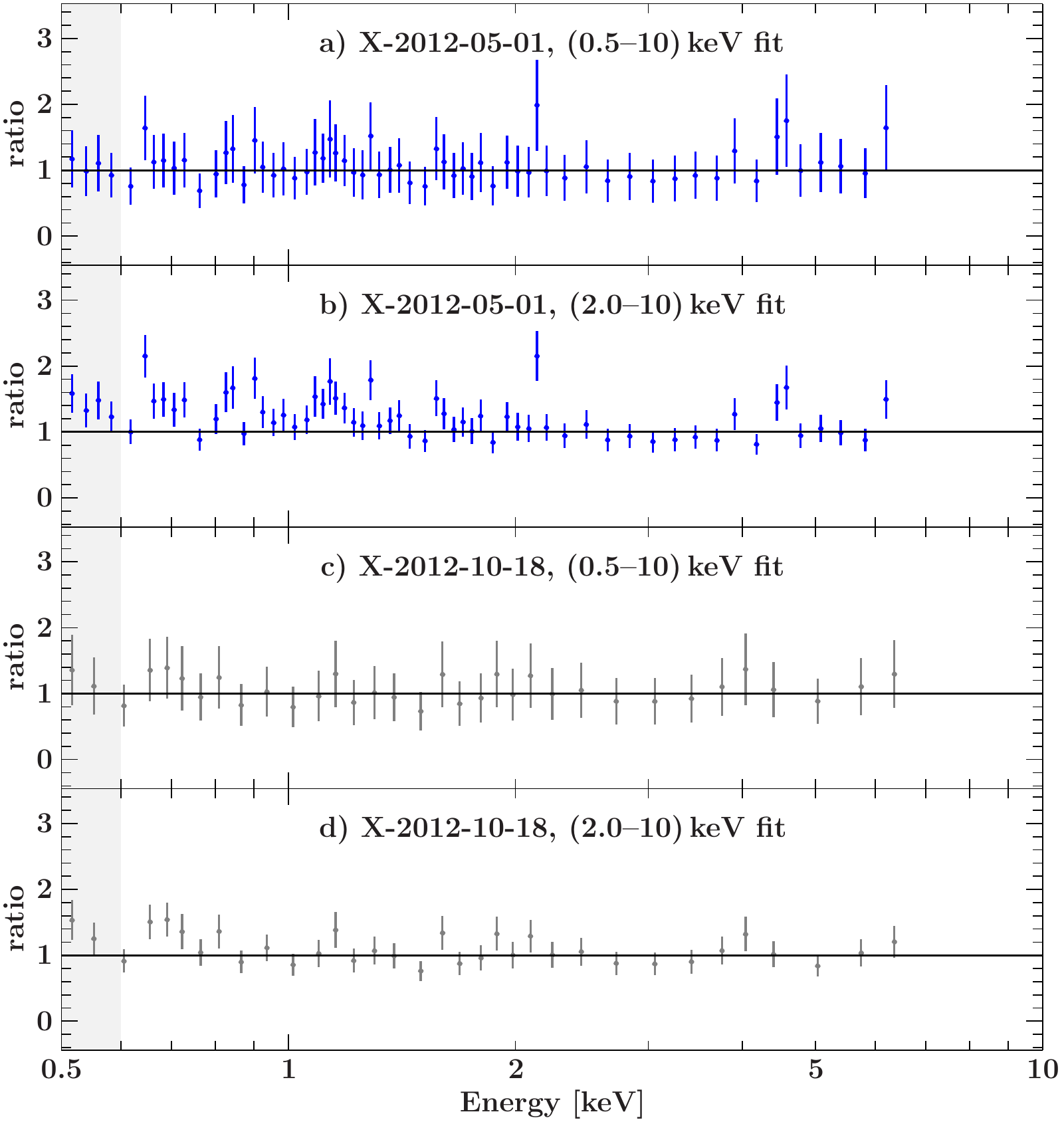}} 
   \caption{ \xmm{} ratio of the \xmm{} EPIC PN difference spectra to a power-law model. Difference spectra  were created by subtracting the background-subtracted spectra of X-2012-05-01 (blue) and X-2012-10-18 (black) from the one of X-2004-04-11. Simple power-law fits were performed on the energy ranges (0.5--7)\,keV and (2.0--7)\,keV, where the fit to the latter band was then extrapolated to lower energies.} 
\label{fig-xmm_diffspec}
\end{figure}

Visual inspection of the \xmm{} data revealed a flat and smooth spectrum for each observation (see Figs.~\ref{fig-xmm_abspow02} and~\ref{fig-xmm_abspow06}). The source is detected from 0.5\,keV to 10\,keV. We first adopted an unbroken power law modulated by Galactic \ion{H}{}\textsc{I} absorption $N_{\rm H,Gal}=3.17\times 10^{20}\,\mathrm{cm}^{-2}$, based on the Galactic Hydrogen LAB survey \citep{Kalberla2005}. The high-resolution ISM absorption model \textsf{tbnew}\footnote{Wilms, J., Jutt,~A.~M., Schulz,~N.~S., Nowak, ~M.~A. (2012), published at http://pulsar.sternwarte.uni-erlangen.de/wilms/research/tbabs/} \citep{Wilms2000} is used to account for neutral absorption. Cross-sections and abundances were taken from \citet{Verner1996} and \citet{Wilms2000}, respectively. The large number of spectral data counts allowed for the use of $\chi^2$-statistics, and each spectrum was grouped to a S/N of five per energy bin. For each observation, we fit the available PN and MOS data in the (0.5--10)\,keV energy range simultaneously. 

The absorbed power-law model yielded a good fit to each spectrum. Best-fit results for the photon index $\Gamma$ lay between 1.50 and 1.65. We found no evidence of intrinsic absorption when leaving the column density of the absorption free to vary. The upper limit of the column density was $N_{\rm H}\le1.7\,N_{\rm H,Gal}$. It was therefore fixed to $N_{\rm H,Gal}$ to reduce the number of free parameters. In the following, unless stated otherwise, all spectral models mentioned in this paper include the Galactic absorption component with fixed column density. The \xmm{} spectra are shown in \mbox{Figs.~\ref{fig-xmm_abspow02} and~\ref{fig-xmm_abspow06},} along with their best-fit models (solid line). We convolved the power-law component of the best-fit model with \textsf{cflux} to determine the unabsorbed energy flux in different energy bands, i.e. total band (0.5--10)\,keV, soft band (0.5--2)\,keV, and hard band (2--10)\,keV. For each spectrum, we performed three independent fits by taking only data within each respective band into account. The sub-bands were defined based on typical flux ranges used in the literature to facilitate comparison. The best-fit parameters and absorption-corrected fluxes of the fit are listed in \mbox{Table \ref{tab-bfpar-pow}}. 
\\
We found no indications of any spectral line emission features. The addition of a Gaussian line at 6.4\,keV with a fixed width of 1\,eV did not yield any significant improvement of the fit (\mbox{$\Delta\chi^2=-0.1$} for one additional free parameter). We measured an upper limit for the equivalent width of $EW_{6.4\kev}\le60\,\mathrm{eV}$ in all three observations. This is consistent with the results of G06.   
\\
G06 also report indications of a weak soft excess below \mbox{$\sim1\,\kev$}. They found that a broken power law with a break energy at $\sim0.6\,\kev$ is a good approximation of the data. Thus, we looked for the existence of such a soft excess by modeling the spectra only above $2\,\kev$  and extrapolating the best fit down to 0.5\,keV. No significant excess emission is observed in the case of X-2012-05-01 or X-2012-10-18, but visual inspection of the residuals suggests a possible soft excess in the data of X-2004-04-11 (see \mbox{Figs.~\ref{fig-xmm_abspow02} and~\ref{fig-xmm_abspow06}}). The spectra of all three \xmm{} observations were then modeled phenomenologically with a broken power law. We note that results for the soft photon index $\Gamma_1$ are not well constrained because of the narrow energy range from 0.5\,keV to the break energy. Fit results for the broken power law are summarized in Table \ref{tab-bfpar-bknpow}. For X-2004-04-11, the model provided a slightly better fit than the simple power law. The best-fit parameters are consistent with results by G06, except for the soft photon index $\Gamma_1$, which cannot be constrained well. As for X-2012-05-01, it was not possible to constrain either of the photon indices when we left the break energy free to vary. We fixed the break energy to 0.6\,keV as reported in G06, which allowed us to obtain constraints on $\Gamma_1$ and $\Gamma_2$. The fit of X-2012-10-18 yields good constraints, however the break energy is significantly higher than in G06. Within its uncertainties, the hard photon index $\Gamma_2$ is roughly consistent with the one from the simple power law in all three observations. 

We tested the significance of the broken power-law model by applying the F test. The null-hypothesis model assumes an unbroken power law as an underlying model. A broken power-law model was considered to be significantly different from the simple power-law model if the probability that the latter is correct was less than 1\% (expressed by the $p$ value). The F-test $p$ values are also listed in Table \ref{tab-bfpar-bknpow}. For X-2012-05-01 and X-2012-10-18, both models were statistically equal. For X-2004-04-11, the broken power law gave a slightly better fit than the simple power law, and the $p$-value was inconsistent with the null hypothesis according to our threshold. Thus, the presence of a weak soft excess is tentative. However, it was striking that no excess was detected in the observations in 2012. Since background count rates were consistent with each other in all three observations, a contribution of systematic background variations to the soft X-rays could be excluded. 

Assuming the soft excess of X-2004-04-11 is caused by another emission component, the spectrum was modeled phenomenologically using a blackbody in addition to the simple power law. The power law was fitted in the (2--10)\,keV band and subsequently fixed to the best-fit parameters (see Table \ref{tab-bfpar-bbody}). 
This model yielded a good fit of $\chi^2/\mathrm{d.o.f.}=396.4/398$. The contribution of the blackbody to the (0.5$-$2)\,keV flux is \mbox{$(0.06\pm0.01)\times 10^{-12}\,\mathrm{erg\,cm}^{-2}\,\mathrm{s}^{-1}$}, which makes up 10-20\% of the power-law flux in the soft band. We applied the same model to the data of 2012 (see Table \ref{tab-bfpar-bbody}) where no soft excess has been observed. The fits were acceptable and yielded a contribution of the blackbody to the soft band between 5\% and$20\%$ of the power-law flux in the same band. We note that for these fits, the power law is slightly steeper, which may in principle hide the presence of a faint extra component. 

Two difference spectra were created by subtracting the background-subtracted EPIC PN spectra of \mbox{X-2012-10-18} and \mbox{X-2012-05-01} from the one of \mbox{X-2004-04-11} (see Fig. \ref{fig-xmm_diffspec}). The spectra were rebinned in the same way as the original data. We ignored energy bins larger than 1\,keV. This leads to no spectral bins being regarded above 7\,keV, because the number of spectral data counts of the X-2012-05-01 spectrum decreases rapidly in this range. We fit the difference spectra by applying a Galactically-absorbed simple power law to each of the (0.5$-$7)\,keV and (2$-$7)\,keV bands independently (see Table \ref{tab-bfpar-diff}). In the total range, the best-fit parameters of both difference spectra are consistent with those of X-2012-10-18 and X-2012-05-01 within 90\% uncertainties. The best-fit values of the photon indices are slightly flatter. This is expected if the spectrum of X-2004-04-11 is harder than in 2012. Fitting only the hard band and extrapolating the fit into the soft band, a soft excess is not observed in the difference spectra, as illustrated in Fig. \ref{fig-xmm_diffspec}.
 
\subsection{\swift{}/XRT}
\label{sec-ana-xrt}
The \swift{} XRT spectra have significantly lower S/N, and in most observations the source is not detected above 6\,keV. Because of the poor quality of the data, we did not adopt a more complex model than the simple power law, which already gave a good description of the \xmm{} data. The model is fitted to the available unbinned data in the (0.5--10)\,keV range using the unbinned likelihood statistic \textsf{Cstat} \citep{Cash}. Best-fit parameters are listed in Table \ref{tab-bfpar-pow}. The low number of spectral data counts mean that the fit parameters were affected by high uncertainties, especially during times of low flux. To improve statistics of the spectral fit, we defined three flux states based on the results of the single observations and merged spectra with fluxes within these intervals (in units of $10^{-12}\,\mathrm{erg\,cm}^{-2}\,\mathrm{s}^{-1}$): \mbox{$0.4< F_{0.5-2\kev}\le 0.9$} (low), \mbox{$0.9< F_{0.5-2\kev}\le 1.4$} (medium),  and \mbox{$1.4< F_{0.5-2\kev}$} (high). This gave a total exposure of 57.02\,ks with 837 counts in the low-flux state, 46.9\,ks with 996 counts in the medium-flux state, and 25.11\,ks with 795 counts in the high-flux state. The quality of the merged spectra then enabled the use of $\chi^2$-statistics and the spectra were analyzed analogously to the \xmm{} spectra. For each of the three spectra, simple power-law models yield good fits with no significant evidence of any other feature such as a soft excess. Best-fit parameters for these fits are listed in Table \ref{tab-bfpar-pow}, labeled as S-low, S-medium, and S-high. There was no evidence of flux-dependent spectral variations.

\renewcommand{\arraystretch}{1.3}

\begin{table*}
  \caption{Results of the absorbed broken power-law model for \xmm{} spectra. }
  \label{tab-bfpar-bknpow}
  \centering
  \begin{tabular}{llllrr}
    \hline \hline
    Inst-Obsdate  &     $\Gamma_1$ & $\Gamma_2$& $E_\mathrm{B}$ & $\chi^2$/d.o.f. & $p$-value\\    
    (1) & (2) & (3) & (4) & (5) & (6) \\
    \hline
    X-2012-10-18 & $1.38^{+0.16}_{-0.17}$ & $1.65^{+0.01}_{-0.01}$ & $1.02^{+0.32}_{-0.16}$ & 288.6/256 & 0.033 \\
    X-2012-05-01 & $2.40^{+1.60}_{-2.10}$ & $1.56^{+0.06}_{-0.06}$ & $0.61$ (fixed) & 134.5/130 & 0.381 \\
    X-2004-04-11 & $\ge 3.05$ & $1.50^{+0.03}_{-0.03}$ & $0.56^{+0.05}_{-0.05}$ & 372.1/396 & 0.0002 \\
    \hline
    \hline
  \end{tabular}
  \tablefoot{ Best-fit parameters for the absorbed broken power law. The absorption is fixed to its Galactic value, $3.17\times 10^{20}\,\mathrm{cm}^{-2}$ \citep{Kalberla2005}. (1) Instrument and observation date.  (2) Soft photon index below $E_\mathrm{B}$. (3)  Hard photon index above $E_\mathrm{B}$. (4) Break energy in units of keV. (5) Fit statistics of the broken power-law model. (6) F-test $p$ value, i.e. likelihood that the null-hypothesis model is correct (see Sect. \ref{sec-ana-xmm} for details). }
\end{table*}

\begin{table*}
  \caption{Results of the black body and power-law model for \xmm{} spectra. }
  \label{tab-bfpar-bbody}
  \centering
  \begin{tabular}{llllllrr}
    \hline \hline
    Inst-Obsdate  & $\Gamma_{2-10\,\mathrm{keV}}$ & $kT$ & $F_\mathrm{BB}$ & $F_\mathrm{PL}$ & $\chi^2$/d.o.f. \\    
    (1) & (2) & (3) & (4) & (5) & (6)  \\
    \hline
    X-2012-10-18 & $1.57\pm0.10$ & $0.31^{+0.10}_{-0.12}$ & $0.012\pm0.005$ & $0.216\pm0.005$ & 291.9/258 \\
    X-2012-05-01 & $1.43\pm0.19$ & $0.23^{+0.06}_{-0.07}$ & $0.023\pm0.005$ & $0.128\pm0.004$ & 134.3/131 \\
    X-2004-04-11 & $1.39\pm0.08$ & $0.25^{+0.06}_{-0.07}$ & $0.059\pm0.009$ & $0.383\pm0.006$ & 396.4/398 \\
    \hline
    \hline
  \end{tabular}
  \tablefoot{ Best-fit parameters for the absorbed simple power law and black body. The absorption is fixed to its Galactic value, $3.17\times 10^{20}\,\mathrm{cm}^{-2}$ \citep{Kalberla2005}. (1) Instrument and observation date. (2) Photon index, fitted in the (2$-$10)\,keV band and fixed for broad band analysis. (3) Black body temperature in keV. (4,5) Absorption-corrected (0.5$-$2)\,keV flux in units of $10^{-12}\,\mathrm{erg\,cm}^{-2}\,\mathrm{s}^{-1}$ of the black body and the power law, respectively. (6) Fit statistics $\chi^2$ per degrees of freedom.}
\end{table*}

\begin{table}
  \caption{Power-law fit of the \xmm{} EPIC PN difference spectra.}
  \label{tab-bfpar-diff}
  \centering
  \begin{tabular}{llrrr}
    \hline \hline
    Subtracted spectrum  & $\Gamma_{2-10\,\mathrm{keV}}$ & $\Gamma_{0.5-10\,\mathrm{keV}}$& $\chi^2$/d.o.f. \\    
    \hline
    X-2012-10-18 & $1.36\pm0.26$ & $1.43\pm0.09$ & 29.4/33\\
    X-2012-05-01 & $1.23\pm0.18$ & $1.45\pm0.06$ & 103.6/69\\
    \hline
    \hline
  \end{tabular}
  \tablefoot{ Best-fit parameters for the simple power-law fit in the (0.5$-$10)\,keV and (2.0$-$10)\,keV energy band to the difference spectra. A difference spectrum is computed by subtracting the respective spectrum from X-2004-04-11. }
\end{table}

\subsection{Multitimescale X-ray variability}
\label{sec-ana-res}
\begin{figure*}
\centering
\resizebox{\hsize}{!}{\includegraphics[width=\linewidth]{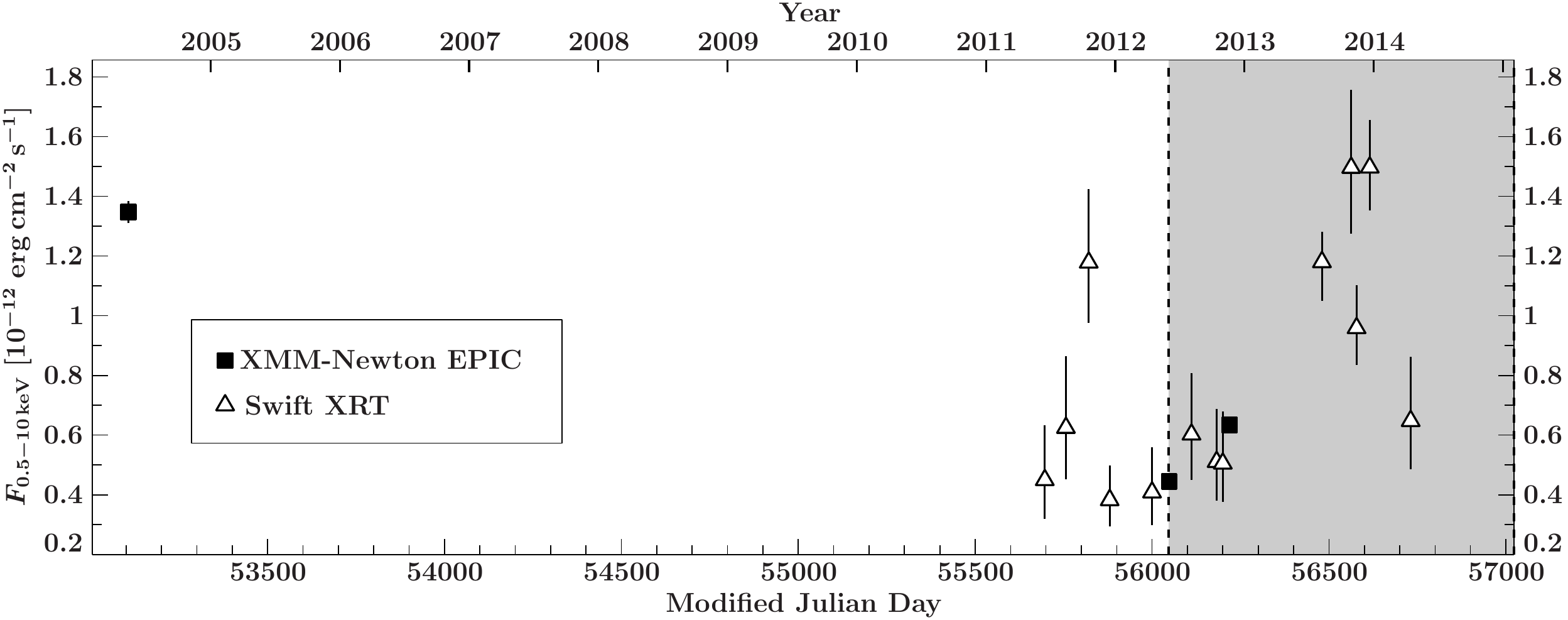}}
\caption{Long-term behavior of the total absorption-corrected 0.5--10\,keV flux. The gray-shaded background denotes the multiwavelength monitoring period. \xmm{} data are shown as black squares, \swift{} data are plotted as triangles.}
\label{fig-lc}
\end{figure*}

In the following, reported fluxes and luminosities refer to their absorption-corrected values. If not stated otherwise, source count rates refer to their background-subtracted values. Figure \ref{fig-lc} shows the evolution of the total flux of the source over time. 

\pks{} shows moderate X-ray flux variability on timescales of weeks to years. The scheduling of the monitoring observations does not allow for detecting variability on timescales shorter than two weeks. For the \xmm{} observations from 2004, the total flux is about 50\% lower in those from 2012 and regained a similar value in September 2013. 

We calculated the fractional variability amplitude $A_{\rm var}$ following the definition by \cite{Vaughan2003}:
\begin{eqnarray}
  A_\mathrm{var}&=&\sqrt{\frac{S^2-\overline{\sigma}^2_\mathrm{err}}{\overline{x}^2}}\\
 \mathrm{err}\left(A_\mathrm{var}\right)&=&\sqrt{\left(\sqrt{\frac{1}{2\,N}}\frac{\overline{\sigma^2_\mathrm{err}}}{\overline{x}^2\,A_\mathrm{var}}\right)^2+\left(\sqrt{\frac{\overline{\sigma^2_\mathrm{err}}}{N}}\,\frac{1}{\overline{x}}\right)^2}
\end{eqnarray}

where $N$ is the number of data points in the time series with a mean of the fluxes $\overline{x}$. The variance and mean square error of the time series are denoted as $S^2$ and $\overline{\sigma^2_\mathrm{err}}$, respectively. 

For the full (0.5--10)\,keV energy range the fractional variability is \mbox{$A^{\rm total}_{\rm var}=0.44\pm0.19$}. If divided into the soft (0.5--2)\,keV and hard (2--10)\,keV bands, we find \mbox{$A^{\rm soft}_{\rm var}=0.39\pm0.18$} and \mbox{$A^{\rm hard}_{\rm var}=0.35\pm0.33$}, consistent with the fractional variability of the total range. As a result, both the soft and hard ranges show the same variability characteristics. 

\begin{figure*}
        \centering
        \begin{minipage}{0.5\linewidth}
                 \resizebox{\hsize}{!}{\includegraphics[width=\linewidth]{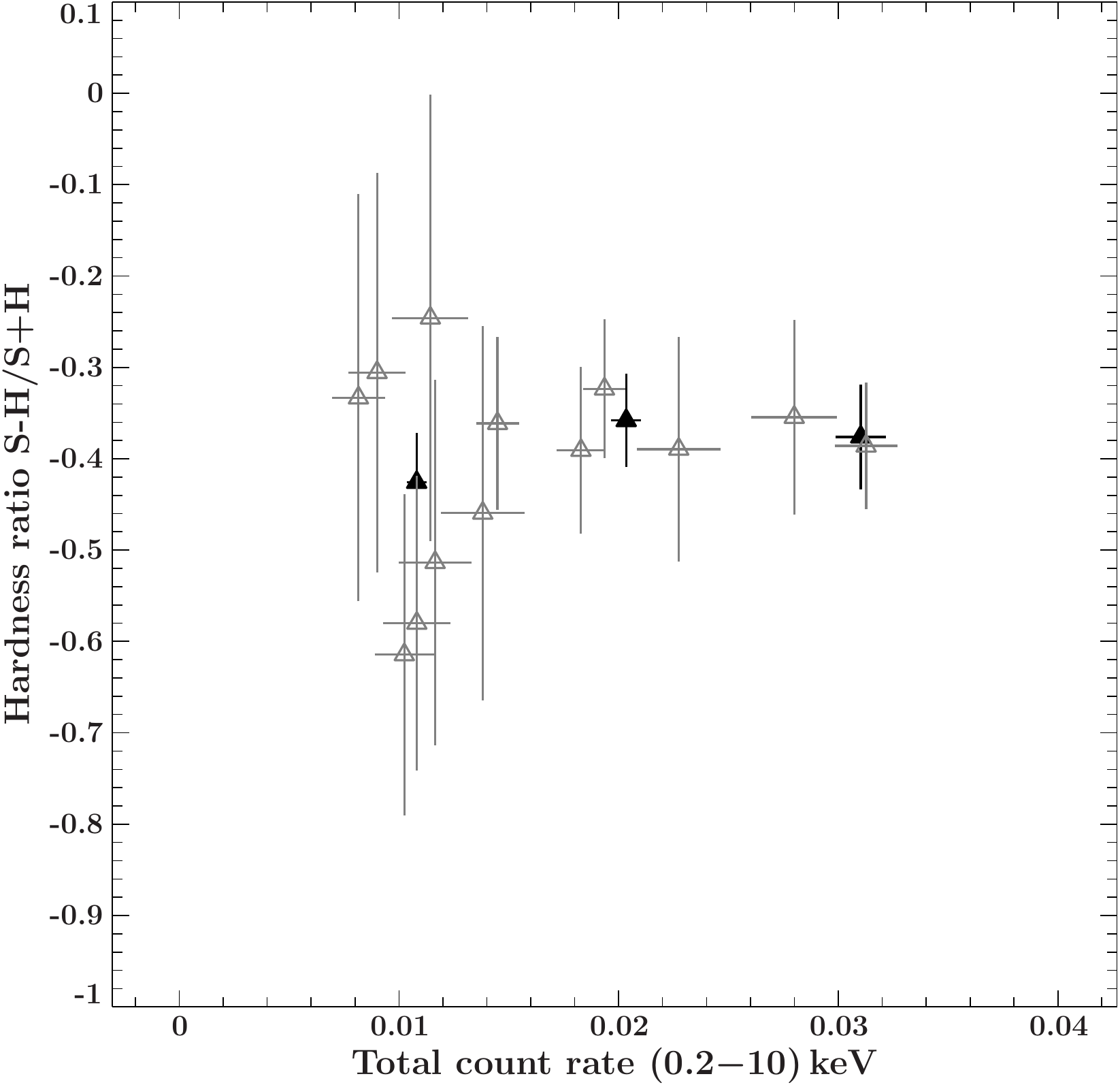}}
         \end{minipage}\begin{minipage}{0.5\linewidth}
        \resizebox{\hsize}{!}{\includegraphics[width=\linewidth]{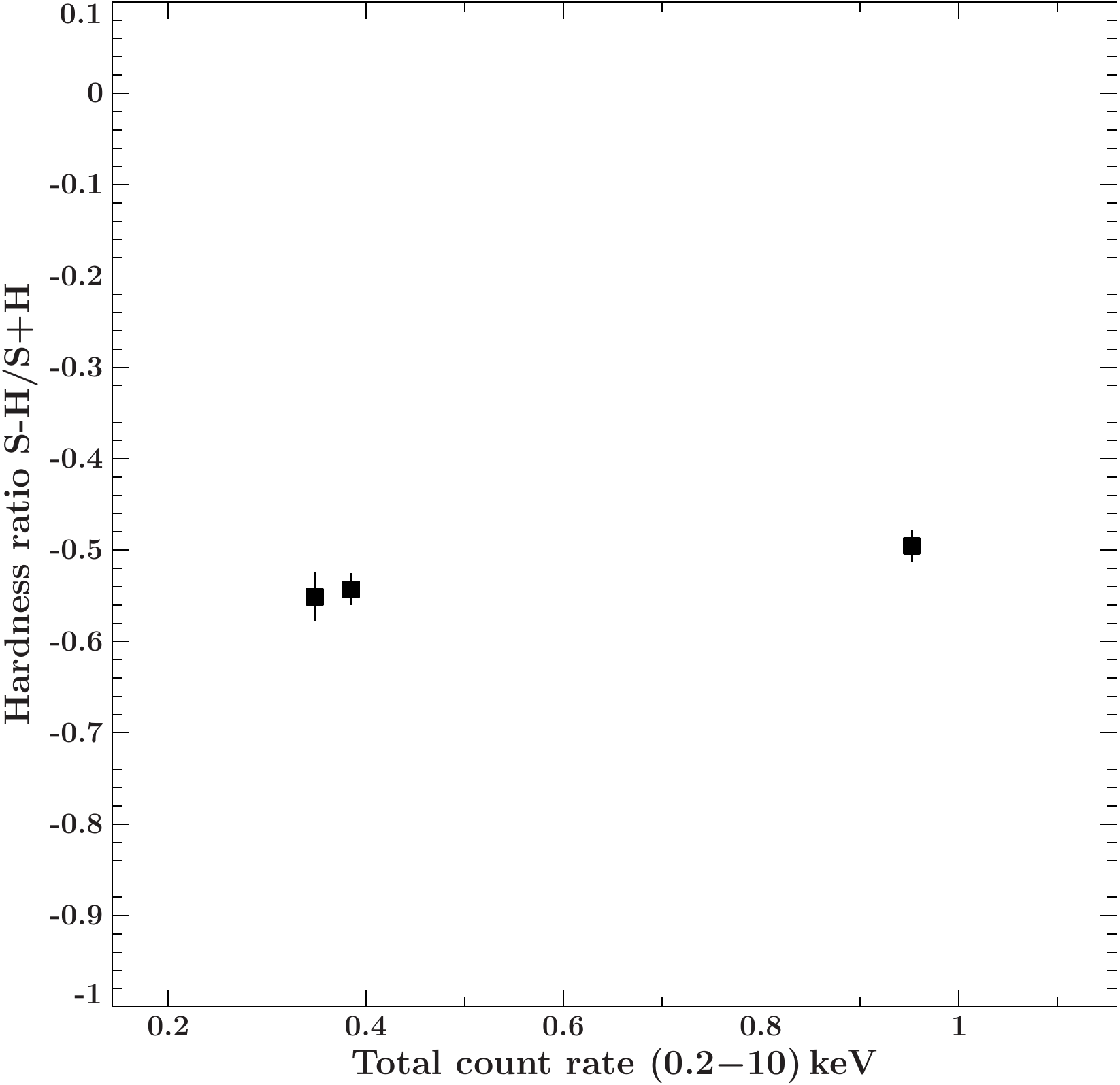}}
        \end{minipage}
        \caption{Hardness-intensity diagram (HID): Hardness ratio S-H/S+H as a function of total background subtracted source count rate. Error bars correspond to 1-sigma uncertainties. \textit{Left:} Results from \swift{} observations. Individual observations are shown as gray open symbols, while filled symbols denote results from merged data sets. \textit{Right:} HID of the \xmm{} data.}
        \label{fig-hid}
\end{figure*}

\begin{figure*}
        \centering
        \begin{minipage}{0.5\linewidth}
        \resizebox{\hsize}{!}{\includegraphics[width=\linewidth]{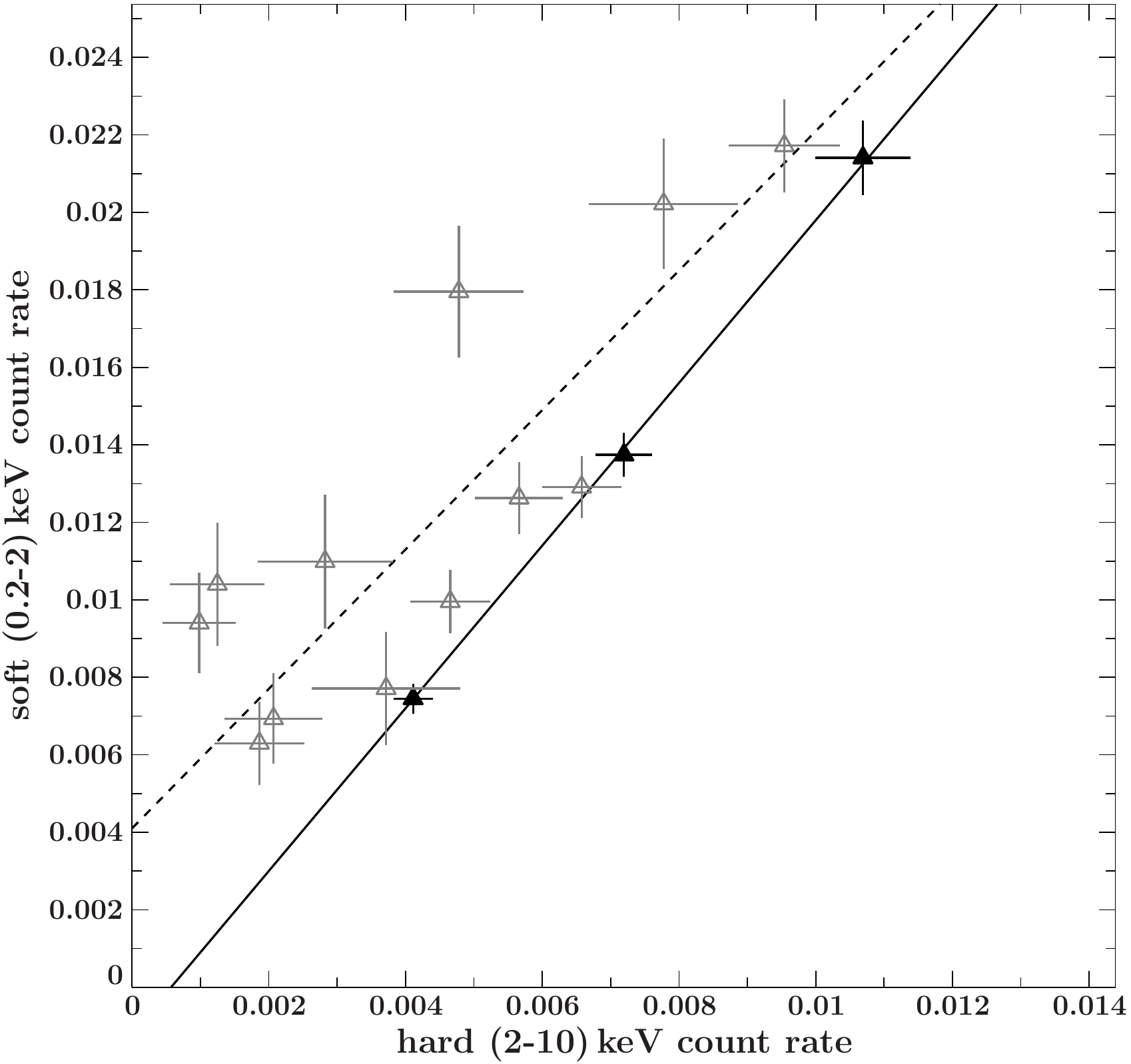}}
 \end{minipage}\begin{minipage}{0.5\linewidth}
        \resizebox{\hsize}{!}{\includegraphics[width=\linewidth]{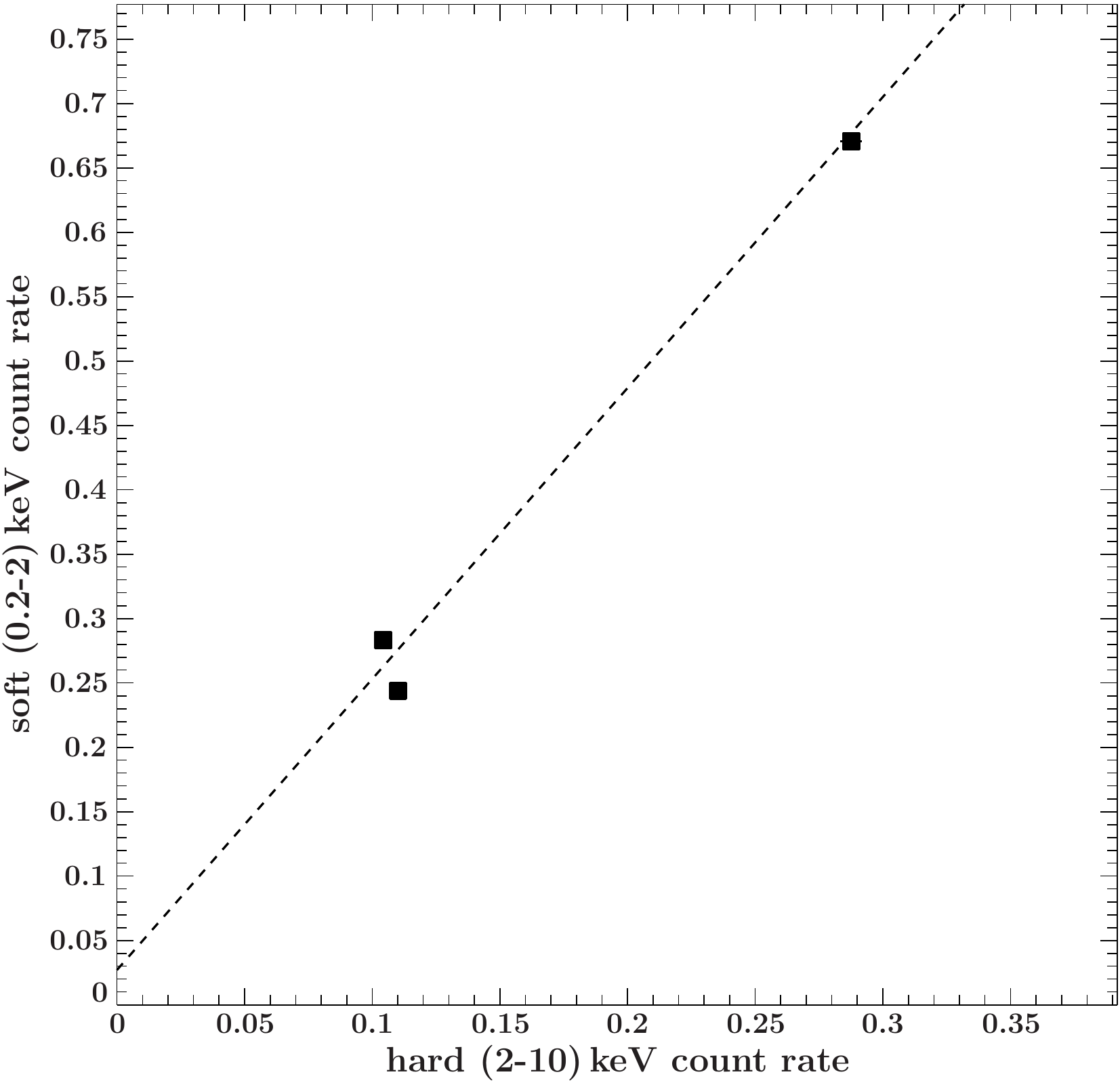}}
        \end{minipage}
        \caption{ \textit{Left:} Hard- versus soft-band diagram (HSD): Comparison of the count rates in the soft (0.2--2)\,keV and hard (2--10)\,keV band. Error bars correspond to 1-sigma uncertainties. Lines denote the best fit of the linear regression for low- (dashed) and high-S/N (solid) data.  \textit{Left:} Results from \swift{} observations. Individual observations are shown as gray open symbols, while filled symbols denote results from merged data sets. \textit{Right:} HSD of the \xmm{} data. Dashed line shows the best fit of the linear regression.}
        \label{fig-hsd}
\end{figure*}

For a model-independent analysis of spectral variability, we determined the number of background-subtracted source counts in the soft and hard bands. Since source counts are not affected by calibrations issues at low energies, we extended the soft band to (0.2--2.0)\,keV. One-sigma uncertainties, $\sigma_N$, of the number of counts, $N$, in a given band were approximated following \citet{Gehrels1986}:
\begin{equation}
 \label{eq_err}
 \sigma_N\approx\sqrt{N+0.75}+1.
\end{equation}
The number of detected counts strongly depends on the energy-dependent effective area of a given instrument. Thus, for similar fluxes we expect different count rates and a number of spectral counts for \swift{} XRT and \xmm{} EPIC PN. There were no simultaneous observations of both \swift{} and \xmm{}, which could have been used to calculate the conversion factor between these instruments. Thus, counts were analyzed separately for each telescope.
\\

Hardness ratios $HR$ were derived using the fractional difference, 
\begin{eqnarray}
 \label{eq_hr}
 HR&=&\frac{S-H}{S+H}, \\
 \sigma_{HR}&=&2\left(S+H\right)^{-2}\sqrt{H^2\sigma^2_S+S^2\sigma^2_H}
\end{eqnarray}
where $S$ and $H$ are source counts in the soft and hard bands with one-sigma uncertainties, $\sigma_S$ and  $\sigma_H$, respectively.

Figure \ref{fig-hid} shows the hardness ratio plotted as a function of the total source count rate (hardness intensity diagram, HID) for the different instruments. For both instruments, the data are consistent with a constant HR. 

Spectral analysis of the \xmm{} data yielded hints of a second spectral component visible below 2\,keV. We tested the presence of a non- or a less-variable component by comparing the variation in the count rates in the soft and hard bands (see Fig. \ref{fig-hsd} for the hard versus soft band diagram, HSD). There is a wide spread at lower count rates. However, the data are overall consistent with a linear relation between the soft and hard bands. 
A linear dependency between these bands suggests that the flux variation of the spectrum is dominated by those of a single component. The presence of a less varying component would induce a positive or negative offset from the origin. We fitted both the whole data set and only high-S/N data with a linear regression and found a relation fully consistent with a line through the origin (see Table \ref{tab-linreg}). The fits were consistent with a single component at 2\,$\sigma$ for all instruments.  

A comparison of background-subtracted count rates of the individual and merged spectra shows that spectra with lower counts appear softer. 

\begin{table}
  \caption{Results from the linear regression to the HSD.}
  \label{tab-linreg}
  \centering
  \begin{tabular}{lrrr}
    \hline \hline
    Data set  & Slope $a$ & Offset $b$ [counts\,s$^{-1}$]\\    
    \hline
    \xmm{}              & $2.26\pm 0.04$ & $  0.027\pm 0.006$ \\
    \swift{}            & $1.84\pm 0.33$ & $  0.004\pm 0.002$ \\
    \swift{} merged     & $2.10\pm 0.43$ & $-0.001\pm 0.002$ \\
    \hline
    \hline
  \end{tabular}
  \tablefoot{ Best-fit parameters for the linear regression  $C_\mathrm{soft}=a C_\mathrm{hard} + b$, where $C_\mathrm{soft}$ and $C_\mathrm{hard}$ are the spectral counts in the soft and hard bands, respectively. Uncertainties correspond to 1-sigma uncertainties.}
\end{table}

\section{Discussion}
\label{sec-ana-dis}
\subsection{The X-ray spectrum of \pks{}}
The X-ray emission of \pks{} in the (0.5$-$10\,keV) energy range is described best by a simple, Galactically-absorbed power law. A tentative soft excess is observed in the \xmm{} spectrum of 2004. However, this excess is not detected in our new observations from 2012. The difference spectrum is described well by a simple power law, but the S/N is not sufficient to draw any conclusion about the tentative excess (see Fig. \ref{fig-xmm_diffspec} and Table \ref{tab-bfpar-diff}). The power-law spectrum and photon index are reminiscent of the non-thermal X-ray emission of type-1 AGN. The spectrum exhibits flux variability across the entire band. Comparing best-fit models of each observation, we find evidence of weak spectral variability, while there is no evidence of any intrinsic absorption. This suggests that the power-law component itself varies, caused by changes in the emission region.  Although we do see evidence of weak spectral variability from spectral fitting, the variations between the source brightness and spectral shape are not correlated (Fig. \ref{fig-hid}). The linear behavior of the count rates in the HSD (Fig. \ref{fig-hsd}) can be explained by a dominating power-law component that exhibits changes in its normalization. These variations cause a homogeneous flux variation across the (0.5--10)\,keV band.  
For interpreting the X-ray spectrum, we consider the following three scenarios: 

(I) a non-jet-dominated X-ray spectrum (characteristic of Seyfert galaxies), for which the unbroken power-law spectrum is associated with thermal inverse-Compton scattering of optical and UV photons from the accretion disk in the disk corona. A soft excess in these sources may be caused by relativistically blurred reflection from the accretion disk \citep{Crummy2006} or thermal Comptonization of disk emission by a population of electrons with low temperature and high optical depth \citep[see, e.g.,][]{HaardtMaraschi1993}. We note that the photon index of \pks{} is significantly harder than those observed in radio-quiet NLS1s \citep[see, e.g.,][]{Vaughan1999,ZhouZhang2010,Grupe2010}.

(II) a Seyfert-like X-ray spectrum that may be contaminated by a significant contribution from a relativistic jet. In this scenario, jet-synchrotron emission contributes to the soft X-ray spectrum in 2004, while the unbroken power law can be associated with the Seyfert-like component. The detection in $\gamma$-rays, however, suggests that the jet dominates at higher energies. Thus, we do not favor any scenario where the jet emission only contributes in the form of a soft excess. 

(III) a jet-dominated X-ray spectrum, characteristic of blazars, in which the spectrum is associated with X-ray emission from the jet alone. Emission from the accretion disk is negligible. Within the framework of leptonic models, blazar X-ray spectra may be comprised of both synchrotron emission and inverse-Compton emission from relativistic particles in the jet plasma \citep[see, e.g.,][]{Maraschi1992, Dermer1993,Bloom1996,Ghisellini1998, Ghisellini2010,Boettcher2013}. In this scenario, the hard power law can be interpreted as inverse-Compton emission from the jet and synchrotron emission would account for the tentative soft excess in the data of 2004.

G06 have modeled the SED of \pks{} from the radio band to hard X-ray band using a simple one-zone synchrotron self-Compton (SSC) model. They find that the X-ray spectrum could not be explained by a single SSC component, but an additional weaker emission component of $(1.2\pm0.4)\times 10^{-14}\,\mathrm{erg\,s}^{-1}\mathrm{cm}^{-2}\mathrm{\,keV}^{-1}$ at 2\,keV is required. They argue that the X-ray spectrum is dominated by thermal Comptonization in addition to a weak SSC component that gives rise to the tentative soft excess. \cite{Paliya2013} also modeled the SED including $\gamma$-ray data using a jet-emission model that consisted of an SSC and an external Compton component. A thermal Comptonization component was not needed. They further conclude that the SED of \pks{} resembles the distribution of flat-spectrum radio quasars (FSRQ). For these sources, hard X-ray emission above 2\,keV is interpreted as inverse-Compton radiation, which suggests a non-thermal origin for the X-ray spectrum of \pks{}. 

The jet-like non-thermal origin for the dominating X-ray emission is supported by the flat photon index with respect to radio-quiet NLS1s. This is consistent with values typically found in blazars, particularly FSRQs \citep[see, e.g.,][]{Fan2012,Rivers2013, Dai2003, Sambruna2004}. In Paper II we analyze the TANAMI VLBI data of \pks{} and find a one-sided core-jet radio structure. Together with the X-ray data, this result favors the third scenario and strongly implies that emission from the relativistic jet is dominating the X-ray spectrum and that additional thermal or non-thermal emission from the accretion disk and its corona may only play a role in very low flux states of the source, as indicated by the HSD (see Fig. \ref{fig-hsd}). 

\subsection{Soft excess}
\label{sec-disc-sx}
A faint soft excess is detected only tentatively in the \xmm{} observation \mbox{X-2004-04-11}, consistent with findings by G06. This component could not be verified by our new \xmm{} observations in 2012. Fitting a power law to the \xmm{} data  only in the (2--10)\,keV range (see dashed lines in Figs. \ref{fig-xmm_abspow02} and \ref{fig-xmm_abspow06}) leads to an excess emission of  \mbox{$F_\mathrm{SX}=(6\pm1)\times10^{-14}\,\mathrm{erg\,s}^{-1}\mathrm{cm}^{-2}$} in the (0.5--2)\,keV band (see Sect. \ref{sec-ana-xmm}). 

It remains questionable whether it is appropriate to interpret this as an additional spectral component. A variable soft component may explain the lack of excess emission in the data of 2012. We note that in radio-quiet Seyferts that include a few NLS1s, there have been cases where the soft excess varies independently from and more slowly than the hard power law \citep[][]{Turner2001,Edelson2002,Mehdipour2011,Arevalo2014}, and even seems to disappear from one X-ray observation to the next on timescales of years \citep[][]{MarkowitzReeves2009, Rivers2012}. Such a component should cause deviations from a linear function through the origin in the HSD, which is not observed.

\subsection{Comparison with compact steep-spectrum sources}
\label{sec-ana-css}
In our radio analysis (see Paper II) we find evidence that \pks{} belongs to the class of CSS sources. CSS and gigahertz-peaked spectrum (GPS) galaxies are bright radio sources, which are considered to be young radio galaxies based on their peaked radio spectra and compact jet structure \citep[e.g.,][and references therein]{Odea1998,Stanghellini2003,Fanti2011,Randall2011}. The same scenario has been discussed for $\gamma$-NLS1 due to their low black hole masses and high accretion rates \citep[e.g.,][]{Foschini2015} . Besides \pks{}, only one other CSS source has been associated with a $\gamma$-ray counterpart in the 3LAC \citep{Ackermann2015}.  The source thus plays an important role in the study of $\gamma$-NLS1 and CSS sources and a possible connection of these classes. 

The X-ray properties of GPS/CSS quasars have been studied by \citet[][S08 hereafter]{Siemiginowska2008} and \citet{Kunert2014}. In most cases, the linear size of GPS/CSS sources in radio is smaller than the spatial resolution of current X-ray instruments. This means that X-ray emission from the entire complex radio structure (i.e., core, jet, and hot spots) is contained within the extraction region of the X-ray point source. Their X-ray spectra can typically be modeled well by a featureless power law without intrinsic absorption. The power-law indices range from 1.5 and 2.2, with the majority between 1.7 and 2.  Their photon index is typically close to those of radio-quiet quasars, but there are cases where the power law is flat, similar to radio-loud sources. S08 report that for these cases, the X-ray emission can be associated with emission from the radio structure. The X-ray absorption-corrected luminosities of GPS/CSS sources are in the range of ($10^{44}-10^{46})\,\mathrm{erg\,s}^{-1}$. 
The X-ray photon index and unabsorbed luminosity of \pks{} thus agree with properties of low-powered CSS sources, where the radio jet contributes to the X-ray spectrum. 
For some GPS/CSS sources, S08 find more complex spectra with evidence of intrinsic absorption of $\sim\,10^{-21}\,\mathrm{cm}^{-2}$ and/or a soft excess. The (0.5-2)\,keV flux contribution of the latter is on the order of $10^{-14}\,\mathrm{erg\,}\mathrm{cm}^{-2}\,\mathrm{s}^{-1}$. The origin of this emission is not yet well understood. Different theoretical predictions include thermal emission through jet-gas interactions with the interstellar medium. The flux of the tentative soft excess of \pks{} in the soft band is consistent with findings by S08. In the context of CSS sources, this may indicate interaction between the young radio jet and the surrounding medium, but further studies are necessary. 

\subsection{Comparison with other $\gamma$-NLS1s and blazars}
\label{sec-ana-comp}
\cite{Foschini2015} have studied the multiwavelength properties of 43 radio-loud NLS1 galaxies, including the seven $\gamma$-NLS1 known so far. \pks{} is the second closest $\gamma$-NLS1. It is also the radio-loudest and only southern-hemisphere source in the small sample. Besides \pks{}, the sources that have been studied in the most detail so far are \pmn{}, \pkss{}, \h{}, and recently, \sbs{} \citep[][]{Abdo2009a,Abdo2009b,Dammando2012,Dammando2013,Yao2015}.
In the radio band, we observe that the 15\,GHz luminosities span two orders of magnitude, and both flux and spectral variability are observed in all sources (Paper II), though the persistent steep radio spectrum in \pks{} is unique in this sample (Paper II).
In this section, we compare our results for \pks{} with characteristics of these sources. Information on their X-ray properties  is taken from the literature \citep[][]{Zhou2007,Yuan2008,Abdo2009a,Dammando2012,Dammando2013,Dammando2013a,Paliya2013,Rivers2013,Bhattacharyya2014,Yao2015}. Table \ref{tab-comp} provides results of the simplest best-fit model for each source. For comparison, we include results for \pks{} from this analysis averaged over all observations.   

 \begin{figure}
 \centering
   \resizebox{\hsize}{!}{\includegraphics[width=\linewidth]{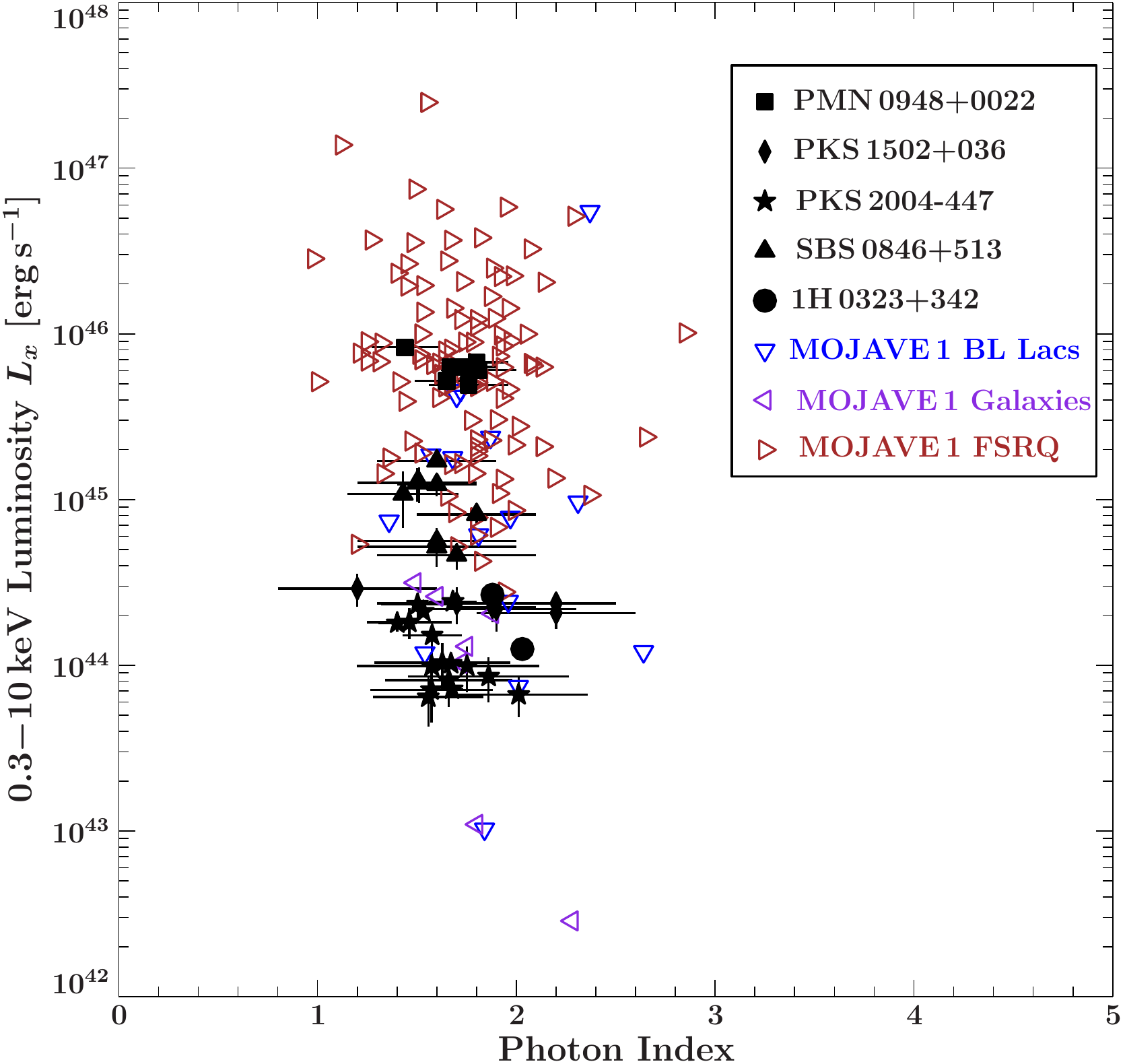}} 
   \caption{X-ray luminosity $L_X$ in the (0.3--10)\,keV band as a function of photon index for $\gamma$-NLS1 galaxies in this paper (black filled symbols) and the MOJAVE\,1 sample (open symbols).}
 \label{fig-lxg}
 \end{figure}
\begin{figure}
\centering
  \resizebox{\hsize}{!}{\includegraphics[width=\linewidth]{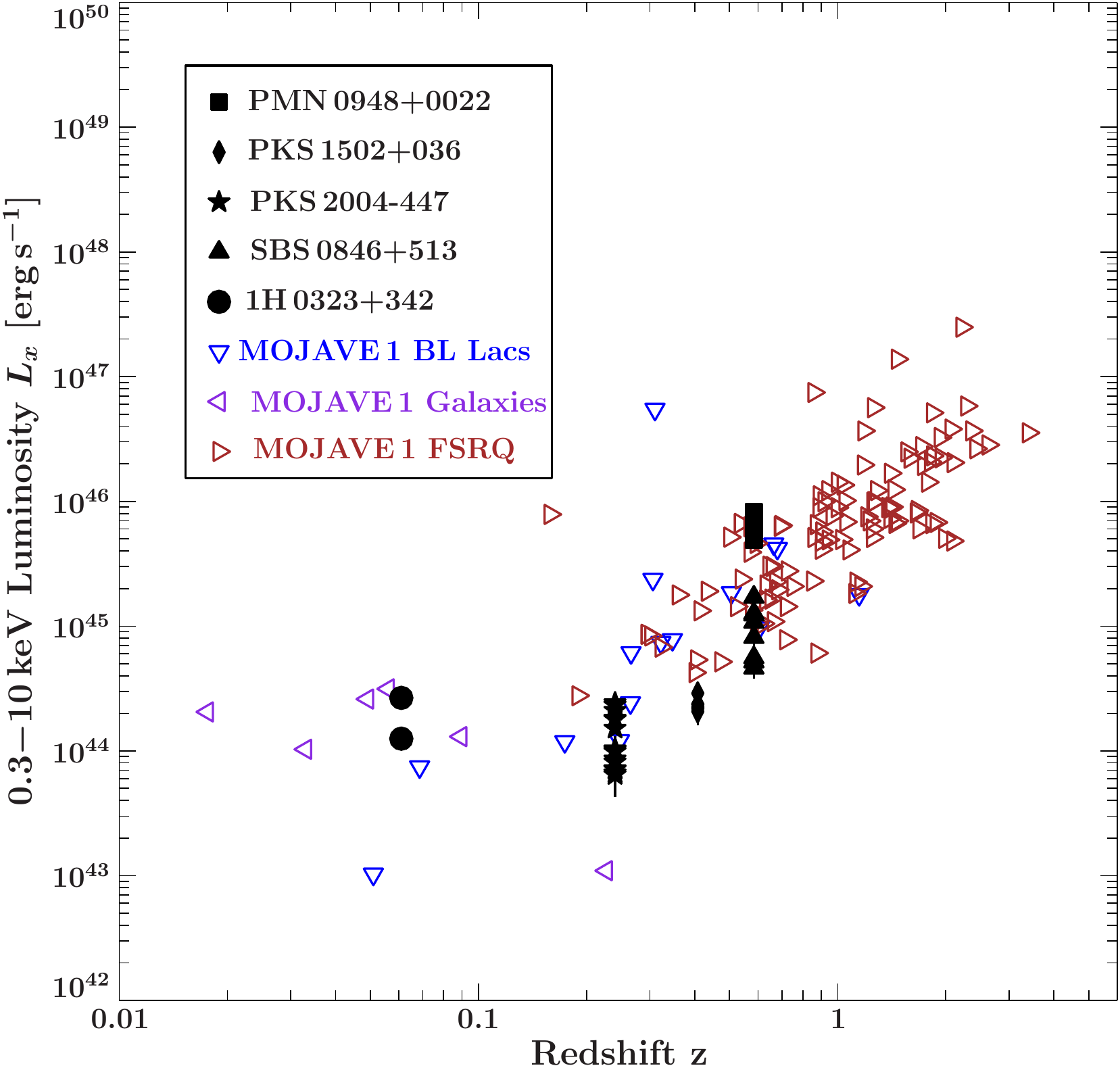}} 
  \caption{Comparison of (0.3--10)\,keV X-ray luminosity $L_X$ as a function of redshift $z$ for the MOJAVE\,1 sample (open symbols) and $\gamma$-NLS1 sources (black filled symbols) discussed in this paper.}
\label{fig-lxz}
\end{figure}
\begin{figure}
\centering
  \resizebox{\hsize}{!}{\includegraphics[width=\linewidth]{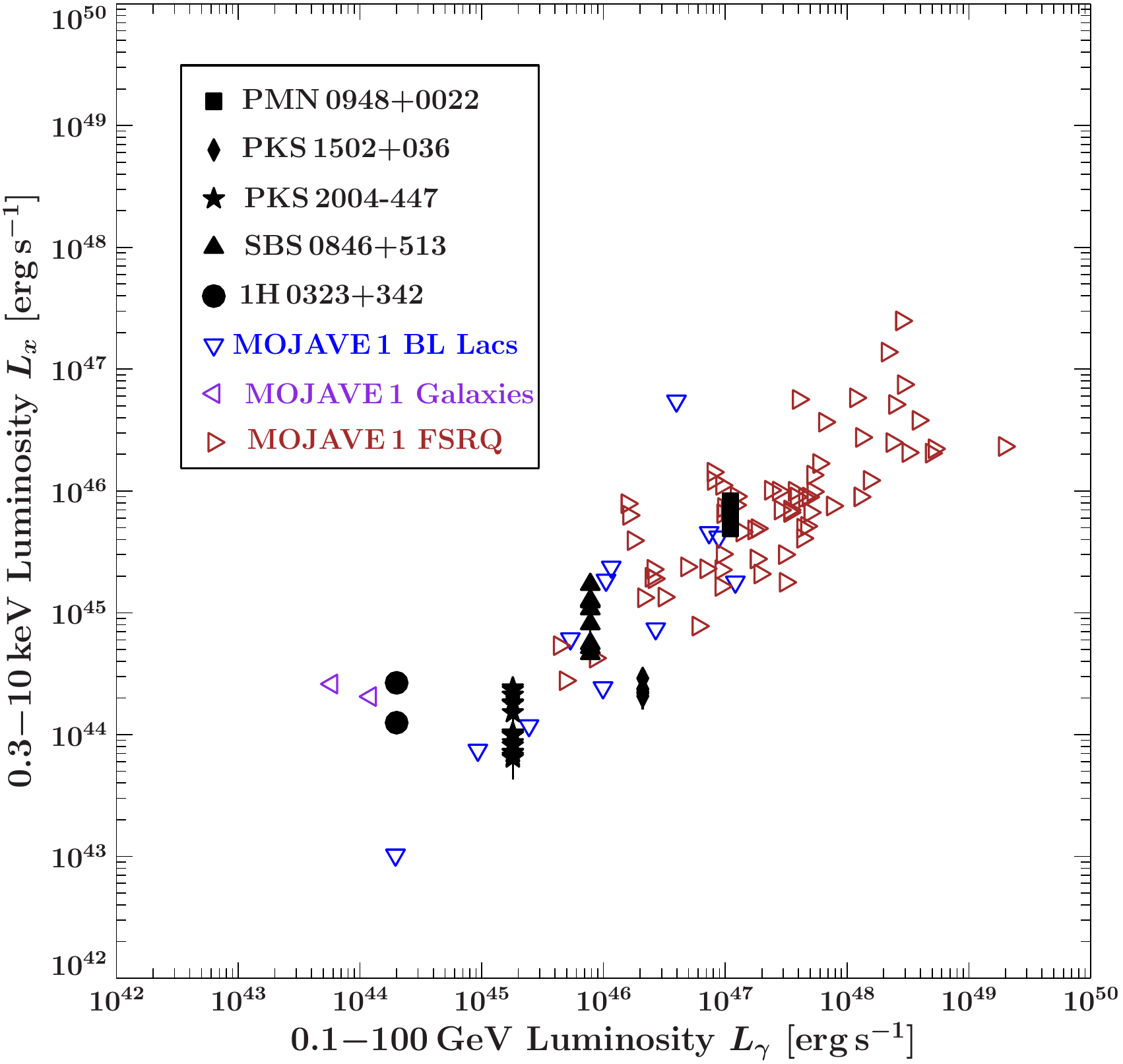}} 
  \caption{Comparison of (0.3--10)\,keV X-ray luminosity $L_X$ as a function (0.1--100)\,GeV luminosity $L_\gamma$ of the  MOJAVE\,1 sample (open symbols) and $\gamma$-NLS1 sources (black filled symbols) discussed in this paper.}
\label{fig-lxlg}
\end{figure}

All five $\gamma$-NLS1 objects show moderate flux variability. Only the closest, $\gamma$-NLS1 ( \h{}), shows flaring activity on timescales of days \citep{Paliya2014,Yao2015}. No evidence was found of  intrinsic absorption in any of these objects. It is therefore reasonable to conclude that variations in flux imply changes in the luminosity and/or spectral shape, rather than changes in the obscuration of the primary radiation. Spectral variability is observed in two cases (\pks{}, \h{}) and suspected in two more (\pkss{}, \pmn{}). 

Results from X-ray analyses reported in the literature mentioned above were used to calculate the X-ray luminosities using the $K$ correction by, for example, \cite{Ghisellini2009}. Figure \ref{fig-lxg} shows the X-ray luminosities of $\gamma$-NLS1s as a function of the photon index. The luminosities span a range of almost two orders of magnitude from $10^{44}\, \mathrm{erg\,s}^{-1}$ to $10^{46} \mathrm{erg\,s}^{-1}$, while the photon indices ranges from 1.2 up to 2.2. However the majority of measurements of $\Gamma$ cluster between 1.5 and 1.8. 

We compare this to the X-ray properties of blazars forming a statistically complete sample of radio-selected extragalactic jets (MOJAVE\,1, \citealt{ListerHoman2005}) that were detected by \fermi{} (see Figs.~\ref{fig-lxg},~\ref{fig-lxz}, and~\ref{fig-lxlg}). The X-ray data on the MOJAVE\,1 sources were obtained from single \swift{} observations \citep{Chang2010}. X-ray luminosities of the $\gamma$-detected MOJAVE\,1 sources span a wider range from $10^{42}\,\mathrm{erg\,s}^{-1}$ to a few $10^{47}\,\mathrm{erg\,s}^{-1}$. In Figs.~\ref{fig-lxz} and~\ref{fig-lxlg}, we show the X-ray luminosity of both samples as a function of the redshift and $\gamma$-ray luminosity taken from the 2FGL catalog \citep{Nolan2012}. For each $\gamma$-NLS1 source, we used the weighted mean of the X-ray luminosities from different observations. The blazar distribution shows the known sequence from high-luminosity FSRQs detected at high redshifts to lower luminosity BL Lac objects observed at lower redshifts \citep[see, e.g.,][]{Fossati1998}. 
The X-ray and $\gamma$-ray luminosities of $\gamma$-NLS1 lie in the same range as blazars. More intriguingly, their luminosities fall into a region that is occupied by BL Lac objects and low-luminosity FSRQs. This result is consistent with the findings of \cite{Paliya2013}, who analyzed non-simultaneous SEDs of two $\gamma$-NLS1s, namely \pks{} and \pkss{}. In comparison to the SED of a typical BL Lac and FSRQ source, they found that their broad-band spectra can be regarded as intermediate between both blazar types, although they bear a closer resemblance to the SED of FSRQs. Similar results have been obtained by independent studies from \citet{Foschini2015} and \citet{Sun2015}.

 \begin{table*}
 \caption{Overview of X-ray spectral properties of $\gamma$-ray emitting and radio-loud NLS1 galaxies.}
 \label{tab-comp}
 \centering
 \begin{tabular}{@{}llllllrcr@{}}
 \hline \hline
 Source & Redshift  &   Model &  $\Gamma_\mathrm{soft}$\tablefootmark{a}     & $\Gamma_\mathrm{hard}$\tablefootmark{b} & $E_\mathrm{B}$\tablefootmark{c} & $F_X$\tablefootmark{d} & Variability\tablefootmark{e} & Ref \\    
 \hline
 \h{}           & 0.061 &  simple power law &-& $2.02\pm0.06$  & - & $14.0\pm0.1$ &F/S & (1)\\
 \pkss{}        & 0.409 &  simple power law &-& $1.7\pm0.2$  & - & $0.3\pm0.7$ & F/(S)&(2)\\
 \pks{}         & 0.240 &  simple power law &-& $1.58\pm0.11$ &-& $0.7\pm 0.4$ & F/S & this paper \\
 \pmn{}         & 0.585 &  broken power law & $2.14^{+0.03}_{-0.02}$ & $1.48^{+0.04}_{-0.03}$ & $1.72^{+0.09}_{-0.11}$ & $4.59^{+0.03}_{-0.05}$ & F/(S) & (3)\\
 \sbs{}         & 0.584 &  simple power law &-& $1.47\pm0.25$  & - & $0.8\pm0.4$ & F & (4)\\
 \hline

 \end{tabular}
 \tablefoot{ Overview of spectral parameters of known $\gamma$-ray and radio-loud NLS1 galaxies for the simplest best-fit models in the literature.
 \tablefoottext{a} {Photon Index, in the case of \pmn{} $\gamma_1$ represents the photon index before the break energy.}
 \tablefoottext{b} {Photon Index after the break energy of the broken power law.}
 \tablefoottext{c} {Break energy in units of keV of the broken power law.}
 \tablefoottext{d} {X-ray flux in units of $10^{-12}\,\mathrm{erg\,s}^{-1}\mathrm{cm}^{-2}$  in the range of (0.3--10)\,keV in case of \pmn{}, \pkss{}, and \sbs{}, and (0.2--10)\,keV for \h{}. }
 \tablefoottext{e} {Variability of the X-ray spectrum. Flux variability is denoted as F, spectral variability as S. Indications of variability that has not been confirmed are shown in parenthesis.} 
\textbf{References:} (1) \cite{Abdo2009a}; (2) \cite{Paliya2013}; (3) \cite{Dammando2013}; (4) \cite{Dammando2012}
}
 \end{table*}

\section{Conclusion}
\label{sec-sum}
As part of the TANAMI multiwavelength program, we analyzed new and archival X-ray observations made by \swift{} and \xmm{} of the only known southern-hemisphere, radio-loud, $\gamma$-ray emitting narrow line Seyfert 1 galaxy \pks{} and compared the results obtained so far for the four other well-studied sources of this type. 

The (0.5--10)\,keV X-ray spectrum of \pks{} is best described by a power law with Galactic absorption. Its photon index is consistent with values typically observed for FSRQs. No evidence for emission or absorption lines has been found. A possible soft excess reported in the literature (G06) could not be confirmed by our new \xmm{} observations in 2012, but its existence in the data set of 2004 cannot be ruled out. Its flux contribution in the soft spectrum of 2004 is less than $\sim20\%$. 

We observed moderate flux changes with a factor of $\sim3$ between the minimum and maximum flux on timescales down to two months. Weak spectral variability was observed on similar timescales. However, the hardness intensity diagram shows no correlated behavior of count rate and hardness ratio. The linear behavior of variations between the (0.5--2)\,keV and (2--10)\,keV bands indicates that no significant brightness-dependent spectral variations occurred. Supported by evidence of a one-sided relativistic jet in the radio data of \pks{} (Paper II; see also \citealt{Orienti2012}), we concluded that the observed spectrum and variability characteristics can be explained by non-thermal emission from the relativistic jet.  

Comparing the X-ray properties of the detected $\gamma$-ray NLS1 galaxies, we found that the major difference between these sources is their luminosity, which spans a range of almost two magnitudes. With the exception of \h{}, all spectra present a flat and a smooth power-law spectrum. A soft excess does not seem to be a frequent spectral component in $\gamma$-NLS1s. Their photon indices are clustered between 1.5 and 1.8. 

Continuum variability is a common property of these sources. It is most prominent in \sbs{}, \pks{}, and \h{}. The analysis of spectral variabilit, however, requires more high-S/N observations. So far, it has only been significantly observed in one out of five sources.
In comparison with typical $\gamma$-ray detected blazars from the MOJAVE\,1 sample, we find that X-ray and $\gamma$-ray luminosities of $\gamma$-NLS1s lie in the range of BL Lac objects. 

In a second paper (Paper II), we presented the radio properties of \pks{} based on VLBI and other radio observations. We compared the radio spectrum and variability with the same $\gamma$-ray NLS1 galaxies as were presented in this paper. While their X-ray spectra show similar steepness and different variability characteristics, the major difference in the radio properties is the radio spectrum. Among the five sources, three show a flat radio spectrum and one shows a GHz peaked synchrotron (GPS) spectrum, while \pks{} is the only steep spectrum source. The luminosities span a broad range of magnitudes similar to those in X-rays. Based on the small linear size of the radio structure and the steep radio spectrum, \pks{} can be considered a CSS source, i.e. a young radio galaxy at a moderately small orientation angle to the line of sight. Its X-ray results are in good agreement with X-ray properties of this source type and thus supports the hypothesis that $\gamma$-NLS1 are young radio sources.  

\begin{acknowledgements}
We acknowledge support by the Bundesministerium f\"ur Wirtschaft und Technologie through the Deutsches Zentrum f\"ur Luft- und Raumfahrt contract 50 OR 1303, the Deutsche Forschungsgemeinschaft under contract WI 1860/10-1, the Spanish MINECO projects AYA2009-13036-C02-02, AYA2012-38491-C02-01, and by the Generalitat Valenciana project PROMETEO/2009/104 and PROMETEOII/2014/057. E.R. and C.S.C. were partially supported by the MP0905 action `Black Holes in a Violent Universe'. C.S.C. was supported by the EU Framework 6 Marie Curie Early Stage Training program under contract number MEST-CT-2005-19669 `ESTRELA'. This research made use of a collection of ISIS scripts provided by the Dr. Karl Remeis observatory, Bamberg, Germany at http://www.sternwarte.uni-erlangen.de/isis/. This research was funded in part by NASA through Fermi Guest Investigator grants NNH10ZDA001N and NNH12ZDA001N (proposal numbers 41213 and 61089). This research was supported by an appointment to the NASA Postdoctoral Program at the Goddard Space Flight Center, administered by Oak Ridge Associated Universities through a contract with NASA. This research makes use of data obtained by \textit{XMM-Newton}, an ESA science mission funded by ESA Member States and the USA (NASA), and \textit{\swift{}}, a NASA mission with international participation.
\end{acknowledgements}

\bibliographystyle{jwaabib} 
\bibliography{bibtex} 
\clearpage
\appendix

\end{document}